\documentclass[a4paper,fleqn,usenatbib]{mnras}
\pdfoutput=1

\usepackage{newtxtext,newtxmath}

\usepackage[T1]{fontenc}
\usepackage{ae,aecompl}

\usepackage{subfigure}
\usepackage{caption} 
\usepackage{fixltx2e}
\usepackage{dblfloatfix}
\usepackage{multirow}
\usepackage{nccmath}
\usepackage{textcomp}
\usepackage{longtable}
\usepackage[dvipsnames,table]{xcolor} 

\usepackage{soul}
\soulregister\cite7
\soulregister\citet7
\soulregister\citep7
\soulregister\ref7
\soulregister\textit7
\soulregister\label7
\soulregister\subsection7

\usepackage{xcolor}

\usepackage{graphicx}	

\title[Correlated orientations of LQG axes]{Correlated orientations of the
  axes of large quasar groups on Gpc scales}

\author[T. Friday et al.]{Tracey Friday,$^{1}$\thanks{E-mail: tjsotherone@hotmail.com}
Roger G. Clowes$^1$
and
Gerard M. Williger$^2$
\\
$^{1}$Jeremiah Horrocks Institute, University of Central Lancashire, Preston
PR1 2HE, UK \\
$^{2}$Department of Physics and Astronomy, University of Louisville,
Louisville KY 40292, USA \\
}

\date{Accepted 2022 Month XX. Received 2022 Month XX; in original form 2021 Month XX}

\pubyear{2022}

\begin{document}
\label{firstpage}
\pagerange{\pageref{firstpage}--\pageref{lastpage}}
\maketitle

\begin{abstract}

Correlated orientations of quasar optical and radio polarisation, and of radio
jets, have been reported on Gpc scales, possibly arising from intrinsic
alignment of spin axes. Optical quasar polarisation appears to be
preferentially either aligned or orthogonal to the host large-scale
structure, specifically large quasar groups (LQGs). Using a sample of 71 LQGs
at redshifts $1.0 \leq z \leq 1.8$, we investigate whether LQGs themselves
exhibit correlated orientation. We find that LQG position angles (PAs) are
unlikely to be drawn from a uniform distribution ($p$-values $0.008 \lesssim
p \lesssim 0.07$). The LQG PA distribution is bimodal, with median modes at
$\bar{\theta}\sim45\pm2^{\circ}, 136\pm2^{\circ}$, remarkably close to the
mean angles of quasar radio polarisation reported in two regions coincident
with our LQG sample. We quantify the degree of alignment in the PA data, and
find that LQGs are aligned and orthogonal across very large scales. The
maximum significance is $\simeq 0.8\%$ ($2.4\sigma$) at typical angular
(proper) separations of $\sim 30^{\circ}$ (1.6 Gpc). If the LQG orientation
correlation is real, it represents large-scale structure alignment over
scales larger than those predicted by cosmological simulations and at least
an order of magnitude larger than any so far observed, with the exception of
quasar-polarisation / radio-jet alignment. We conclude that LQG
alignment helps explain quasar-polarisation / radio-jet alignment, but raises
challenging questions about the origin of the LQG correlation and the
assumptions of the concordance cosmological model.


\end{abstract}




\section{Introduction}
\label{secIntro}

The spins of galaxies tend to align with the cosmic web of filaments, sheets,
and voids. For example, \citet{Zhang2013} find the major axes of Sloan
Digital Sky Survey \citep[SDSS,][]{York2000} DR7 galaxies are preferentially
aligned with the direction of filaments and within the plane of sheets, and
\citet{Tempel2015} find that orientation of SDSS DR10 galaxy pairs is aligned
with their host filaments. Recently, \citet{Welker2020} detected a
mass-dependent transition of galaxy spin alignments with filaments, from
parallel at low-mass to orthogonal at high-mass. They found that this shift
occurred at $10^{10.4-10.9}$ M\textsubscript{\(\odot\)}, consistent with
Horizon-AGN predictions \citep{Dubois2014,Codis2018}.

Cosmological simulations such as Horizon-AGN \citep{Dubois2014} and
S\textsc{imba} \citep{Dave2019} predict the spin of dark matter haloes (and
galaxies) are preferentially aligned with filaments and sheets at low masses
(mainly spirals) and orthogonal at high masses (mainly ellipticals)
\citep[e.g.][]{Dubois2014,Codis2018,Kraljic2020}. Using the \textit{Planck}
Millennium simulation \citep{Baugh2019}, \citet{GaneshaiahVeena2018}
demonstrate this is a result of accretion history, with low-mass haloes
tending to accrete mass from orthogonal to their host filament and thus
orientating their spins along the filaments. In contrast, they find high-mass
haloes tend to accrete along their host filament and have spins orthogonal to
them.

The spins of quasars are also thought to align with the cosmic
web. Large-scale alignment of quasar polarisation was first reported by
\citet{Hutsemekers1998}, who found the polarisation of optical light from
quasars was coherently oriented on Gpc scales at redshifts of \mbox{$1
  \lesssim z \lesssim 2$}. This was then confirmed at higher significance
levels by further polarisation observations at optical wavelengths
\citep{Hutsemekers2001,Cabanac2005,Hutsemekers2005}, the introduction of
coordinate-invariant statistics by \citet{Jain2004}, analysis using a new and
completely independent statistical method proposed by \citet{Pelgrims2014},
and polarisation measurements at radio wavelengths
\citep{Tiwari2013,Pelgrims2015}. Quasar-polarisation alignment, although
widely and independently reported, remains somewhat controversial
\citep[e.g.][]{Joshi2007,Tiwari2019}.  At least some of the controversy,
however, appears to arise from different authors considering different scales
and using different approaches to test for alignments
\citep[e.g.][]{Pelgrims2015}.

Potential line-of-sight mechanisms for the large-scale alignment of quasar
polarisation must be considered. From the first detection, interstellar
polarisation was a concern but deemed unlikely \citep{Hutsemekers1998}. More
recently, \citet{Pelgrims2019} finds the alignments are robust against
Galactic dust contamination. Another potential line-of-sight mechanism widely
discussed is exotic particles, such as axion-photon mixing in external
magnetic fields
\citep[e.g.][]{Cabanac2005,Das2005,Hutsemekers2005,Payez2008,Agarwal2011,Hutsemekers2011},
although this is disfavoured using constraints from circular polarisation
measurements \citep{Hutsemekers2010,Payez2011}.

If polarisation is not induced along the line-of-sight, we must consider
instrinsic alignment of the quasar spin axes
\citep[e.g.][]{Hutsemekers1998,Cabanac2005,Pelgrims2016thesis}.
\citet{Hutsemekers2014} report that optical quasar polarisation is
preferentially either aligned or orthogonal to the host large-scale
structure. They propose that this bimodality is due to the orientation of the
accretion disk with respect to the line-of-sight, and conclude that quasar
spin axes are likely parallel to their host large-scale
structures. \citet{Pelgrims2016} report a similar result using radio
wavelengths and large quasar groups (LQGs). They also conclude that the
quasar spin axes are preferentially parallel to the LQG major axis for LQGs
with at least 20 members, although they suggest this becomes orthogonal with
fewer members ($10 < m < 20$).

Several studies report that radio jets are aligned over large scales
\citep{Taylor2016,Contigiani2017,Mandarakas2021}, supporting the intrinsic
alignment explanation independently of polarisation measurements. (The work
by \citet{Mandarakas2021} appears to supersede earlier work by the same
group, \citet{Blinov2020}, in which no alignment was found.) The potential
correspondence of the QJARs (quasar jet alignment regions) from
\citet{Mandarakas2021} with other large-scale structures such as the regions
of correlated polarisations is a notable feature. In general, the
corroboration of very large structures by independent tracers can provide
compelling support.

In this paper we investigate for the first time whether LQGs exhibit coherent
orientation, and whether this can explain the reported alignments of quasar
polarisation from \citet{Hutsemekers1998} to \cite{Pelgrims2019}. This
examines scales larger than those so far analysed, and potentially offers
corroborating evidence for, and enhancement of, the intrinsic alignment
interpretation of the results from many quasar polarisation studies
\citep[e.g.][]{Hutsemekers2001,Jain2004,Cabanac2005,Tiwari2013,Pelgrims2014,Pelgrims2015}.
If true, it would represent large-scale structure alignments over $\gtrsim$
Gpc scales, larger than those predicted by cosmological simulations and
larger than any so far observed.

The concordance model is adopted for cosmological calculations, with
$\Omega_{T0} = 1$, $\Omega_{M0} = 0.27$, $\Omega_{\Lambda 0} = 0.73$, and
$H_0 = 70$~kms$^{-1}$Mpc$^{-1}$. All sizes given are proper sizes at the
present epoch.

\section{Data and methods to detect LQGs and measure their orientation}
\label{secLQGDataMethod}

\subsection{Detecting large quasar groups}
\label{subLQGMethods}

Our LQG sample is taken from the work of \citet{Clowes2012,Clowes2013}. The
LQGs were detected using quasars from the Sloan Digital Sky Survey
\citep[SDSS,][]{York2000}, specifically Quasar Redshift Survey Data Release 7
\citep[DR7QSO,][]{Schneider2010}. The DR7QSO catalogue of 105,783 quasars
covers a region of $\sim 9,380 \ \mathrm{deg}^2$, with its main contiguous
area of $\sim 7,600 \ \mathrm{deg}^2$ in the north Galactic cap (NGC).

\citet{Clowes2012,Clowes2013} restrict their quasar sample to low-redshift
($z \leq 2$) quasars with apparent magnitude $i \leq 19.1$ in order to
achieve an approximately spatially uniform sample
\citep{VandenBerk2005,Richards2006}. They further restrict their sample to a
redshift range of $1.0 \leq z \leq 1.8$, within which the proper number
density of quasars as a function of redshift is sufficiently flat for
clustering analysis. They then detect LQGs using a three-dimensional
single-linkage hierarchical clustering algorithm, also known as
friends-of-friends (FoF, Appendix~\ref{appLQGfinder}).

The resultant LQG sample\footnote{Clowes (2016), private communication}
contains 398 LQGs. In order to confidently determine the geometric properties
of the LQGs (e.g.\ orientation and morphology) we restrict their original
sample to those with membership $m \geq 20$, giving a sample of 89 LQGs.

We select the most convincing of these using the significance estimates of
\citet{Clowes2012,Clowes2013}. While the absolute values of these may be
contentious \citep{Nadathur2013,Pilipenko2013}, they provide a legitimate
relative order for ranking based on confidence. As a compromise between
sample size and confidence, we restrict our sample to LQGs with
`significance'\footnote{We attribute no significance to the value of 2.8; it
  is used as a relative threshold only} $\geq 2.8 \sigma$, yielding 72 LQGs.

We finally exclude one LQG in the south Galactic cap, giving our final sample
of 71 LQGs, of varied and generally irregular morphologies, as shown in
Appendix~\ref{appLQGsample}.

\subsection{Determining large quasar group orientation}
\label{subLQGOrientations}

The position angle (PA) of a large quasar group can be calculated in either
two or three dimensions. For our sample of 71 LQGs we find that the two
approaches are generally consistent. We use the 2D approach, which involves
tangent plane projection of the LQG quasars, followed by orthogonal distance
regression (ODR) of the projected points. However, data from the 3D approach,
which involves principal component analysis of the LQG quasars' proper
coordinates, are used for some preliminary analysis of the morphology of
LQGs. See Appendix~\ref{appLQGorientation} for details of both approaches.

Due to the filamentary nature of LQGs, orthogonal distance regression gives a
better linear fit for some LQGs than others. Therefore, we have higher
confidence in some PAs than others. We weight the PA of each LQG according to
its ODR goodness-of-fit by inverse residual variance per unit length as

\begin{ceqn}
\begin{equation} \label{eqWeight}
    w = \ell / \sigma^2\ ,
\end{equation}
\end{ceqn}

\noindent where $\ell$ is the length of the ODR line fitted to the LQG, and
$\sigma^2$ is the residual variance of the $m$ quasars in the LQG, calculated
as

\begin{ceqn}
\begin{equation}
    \sigma^2 = \frac{1}{m-1} \sum_{q=1}^{m} e_q^2\ ,
\end{equation}
\end{ceqn}

\noindent where $e_q$ is the orthogonal residual of the $q^{\mathrm{th}}$
quasar from the ODR line. Note that this definition of weight
(Eq.~\ref{eqWeight}) is dimensionless only after normalization. Where
possible we apply our statistical methods (section~\ref{secLQGStatsMethods})
to both unweighted and weighted PAs.

We measure large quasar group orientation as the position angle from
celestial north. It is important to recognise that the PA data are axial
[$0^{\circ}$, $180^{\circ}$), more specifically 2-axial; $0^{\circ}$ and
  $180^{\circ}$ are equivalent. In addition to analysing raw 2-axial PAs,
  some of our statistical methods (section~\ref{secLQGStatsMethods}) require
  these to be transformed to vector (circular) data [$0^{\circ}$,
    $360^{\circ}$). Following \citet{Hutsemekers2014} and
    \citet{Pelgrims2016thesis}, we also test for alignment and,
    simultaneously, for orthogonality, using 4-axial data [$0^{\circ}$,
      $90^{\circ}$). See Appendix~\ref{appAxial} for details of these
      transformations.

We estimate PA measurement uncertainties using bootstrap re-sampling. For
each of our sample of 71 LQGs, we create $n = 10000$ bootstraps and
calculate their PAs. We find that the circular mean of the bootstraps
generally agrees well with the observed PA, with a mean (median) half-width
confidence interval (HWCI) of $\sim10^{\circ}$ ($\sim8^{\circ}$). See
Appendix~\ref{appPAuncertainties} for details of the bootstrap method, and
calculation of HWCIs and their circular means.

\subsection{Coordinate invariance: parallel transport}
\label{subParallelTransport}

Position angles are dependent on the coordinate system in which they are
measured, and in particular the position of the pole used to define $\theta =
0^{\circ}$. To overcome this coordinate dependence we follow studies of
galaxy spin alignment \citep[e.g.][]{Pen2000}, CMB polarisation
\citep[e.g.][]{Challinor2002}, and quasar spin alignment
\citep[e.g.][]{Jain2004} and use parallel transport.

For two objects at locations $P_1$ and $P_2$ on the celestial sphere,
\citet{Jain2004} proposed parallel transporting the vector at the location of
one object to the location of the other before comparing them. The path they
use is the geodesic (great circle) between the objects. Parallel transport
preserves the angle between the PA vector and the vector tangent to this
geodesic. The correction to apply between $P_1$ and $P_2$ is the difference
between the angles the geodesic makes with one of the basis vectors at each
location. That is, if the tangent plane to the sphere has local basis vectors
($\hat{\theta}_1$, $\hat{\phi}_1$) at location $P_1$, and the tangent unit
vector to the geodesic at this point is given by $\hat{t}_1$, then the angle
$\xi_1$ between $\hat{t}_1$ and $\hat{\phi}_1$ is given by
\citep{Pelgrims2016thesis}

\begin{ceqn}
\begin{equation}
    \xi_1 = \tan^{-1}(-\hat{t}_1 \cdot \hat{\theta}_1,\ \hat{t}_1 \cdot \hat{\phi}_1) \ ,
\end{equation}
\end{ceqn}

\noindent with angle $\xi_2$ at location $P_2$ being similarly obtained.

The parallel transport correction between locations $P_1$ and $P_2$,
i.e.\ the angle by which a vector rotates during parallel transport from
$P_1$ to $P_2$, is given by the difference between angles $\xi_1$ and $\xi_2$
\citep{Jain2004}. So, to parallel transport the position angle $\theta_k$ of
object $k$ to the location of object $i$ we compute

\begin{ceqn}
\begin{equation} \label{eqPTCorrections}
\begin{aligned}
    \theta_k^{(i)} &= \theta_k + \Delta_{k \rightarrow i} \ , \\
    &= \theta_k + \xi_k - \xi_i \ ,
\end{aligned}
\end{equation}
\end{ceqn}

\noindent where $\theta$ refers to position angle, not spherical
coordinates. Applying these corrections results in coordinate-invariant
statistics \citep{Jain2004, Hutsemekers2005}. The result of parallel
transporting a vector from $P_1$ to $P_2$ depends on the path taken between
them. If a different path was chosen the parallel transport correction
($\Delta_{k \rightarrow i}$) would differ.

\subsection{Mock LQG catalogues}
\label{subMockLQGs}

To assess compatibility of the observed LQG PA distribution with that
expected in the $\Lambda$CDM cosmological model we use mock LQG catalogues
constructed by \citet{Marinello2016}. They take a snapshot of the Horizon Run
2 (HR2) simulation \citep{Kim2011} at redshift $z = 1.4$, and divide the
volume into 11 sub-volumes. They then create quasar samples by applying a
semi-empirical halo occupation distribution (HOD) model 10 times to each of
the 11 sub-volumes. Finally, they use the LQG finder of
\cite{Clowes2012,Clowes2013} (Appendix~\ref{appLQGfinder}) to construct 110
mock LQG catalogues.

We restrict each mock catalogue to LQGs with membership $m \geq 20$ and
significance $\geq 2.8 \sigma$. The mean number of LQGs in our mocks is
$\bar{n} = 30\pm 0.4$; numbers in individual mocks vary $20\leq n \leq
42$. We stack these to increase the statistical power. The 10 quasar mock
catalogues created from each sub-volume are not truly independent
\citep{Marinello2015thesis}; each HOD model realization samples the same set
of dark matter haloes. Therefore, for each realization we stack the 11
sub-volumes, which are independent. The mean number of LQGs in each stack is
$\bar{n} = 330\pm 3$. The total number of LQGs in all 110 mocks is 3,296.

We calculate position angles for mock LQGs as for our observed sample,
including applying parallel transport corrections.

\section{Methods for the statistical analysis of LQG position angles}
\label{secLQGStatsMethods}

Statistical analysis of large quasar group position angle data requires
appropriate methods. LQGs are widely and non-uniformly distributed, both on
the celestial sphere (separation $\lesssim 120^{\circ}$) and in redshift ($1
\le z \le 1.8$), and their PAs are axial data. Furthermore, the PA
distribution may be bimodal, with PAs both aligned and orthogonal
\citep{Hutsemekers2014,Pelgrims2016thesis}. Many statistical methods lack
discriminatory power in multimodal cases.

We use methods for statistical analysis of the uniformity, bimodality and
correlation of LQG PAs, specifically to determine:

\begin{itemize}
    \item Are they likely to be drawn from a uniform distribution?
    \item Is their distribution bimodal, and where are the peaks?
    \item Are they more correlated than random simulations?
\end{itemize}

\subsection{Uniformity tests}
\label{subUniformityMethods}

For coordinate invariance, we perform uniformity tests on PAs parallel
transported to the centre ($\alpha = 193.6^{\circ}$, $\delta = 24.7^{\circ}$,
J2000) of the A1 region \citep{Hutsemekers1998} of large-scale alignment of
the polarisation of quasars, which is roughly at the centroid of the LQG
distribution. (Alternative centres for parallel transport are discussed in
section~\ref{subsecBimodalityResults}.) We test the PA distribution for
departure from uniformity using Kuiper's test, the Hermans-Rasson (HR) test,
and the $\chi^2$ test.

\subsubsection{Kuiper's test}
\label{subsubKuipers}

Kuiper's test \citep{Kuiper1960} is a rotationally invariant version of the
better-known Kolmogorov–Smirnov (KS) test. It quantifies the maximum positive
\emph{and} negative differences between an empirical cumulative distribution
function (EDF; our PAs) and a theoretical cumulative distribution function
(CDF; in this case uniform).

To incorporate weighting we compute a weighted EDF, where for any measurement
$x$, $F^w_{EDF}(x)$ is equal to the sum of the normalized weights of all
measurements less than or equal to $x$. Following \citet{Monahan2011}, that
is

\begin{ceqn}
\begin{equation} \label{eqKuiperWeighted}
    F^w_{EDF}(x) = \sum_{i=1}^{n_x} w_i \bigg/ \sum_{i=1}^{n} w_i\ ,
\end{equation}
\end{ceqn}

\noindent where $n$ is the total sample size, $n_x$ is the number of
measurements up to and including $x$, and $w_i$ are their goodness-of-fit
weights (Eq.~\ref{eqWeight}). Kuiper's test is then computed normally, using
$F^w_{EDF}(x)$ in place of $F_{EDF}(x)$.

The rotational invariance of Kuiper's test makes it independent of the
`origin' PAs are measured against (in this case celestial north). This makes
Kuiper's test appropriate for circular and axial data that `wrap' between one
end of the distribution and the other, and also gives it equal sensitivity at
all values of $x$.

The $p$-values are evaluated by simulation. We generate 10000 samples of $n$
random PAs, drawn from a uniform distribution, apply the same weighting, and
calculate the fraction of samples with Kuiper's test statistic at least as
extreme as the observations. Note that using alternative weights
(e.g.\ $w_i^2$, $w = 1/\gamma$, $w = 1/\gamma^2$ where $\gamma$ is the
half-width confidence interval) does not significantly affect the $p$-value.

We apply Kuiper's test to both unweighted and weighted PA distributions of
both 2-axial and 4-axial PAs.

\subsubsection{Hermans-Rasson test}
\label{subsubHR}

\citet{Landler2018} test the performance of the Rayleigh and Kuiper’s tests
(amongst others) with a variety of multimodal distributions. They show that
these tests lack statistical power in most multimodal cases, and find the
Hermans-Rasson (HR) test for uniformity on the circle \citep{Hermans1985}
significantly out-competes the alternatives. The HR method is a family of
tests, based on decomposing a circular distribution using Fourier series
\citep{Landler2019}. Variants of the HR test are controlled by the parameter
$\beta$, with $\beta = 2.895$ being recommended by both \citet{Hermans1985}
and \citet{Landler2018}\footnote{Recommendation in electronic supplementary
  material 2} as offering power in both unimodal and multimodal cases. In
this case, the HR statistic $T$ of $n$ measurements $\theta_1,...,\theta_n$
is defined \citep{Landler2018} as

\begin{ceqn}
\begin{equation} \label{eqHR}
    T = \frac{1}{n} \sum_{i=1}^{n} \sum_{j=1}^{n} \pi - \lvert\pi - \lvert\theta_i - \theta_j\rvert\rvert + 2.895\lvert\sin(\theta_i - \theta_j)\rvert\ ,
\end{equation}
\end{ceqn}

\noindent where $\theta_i$ is the PA of the $i^{\mathrm{th}}$ LQG and
$\theta_j$ is the PA of the $j^{\mathrm{th}}$ LQG, from a sample of $n$ LQGs.

The $p$-values are evaluated by simulation. We generate 10000 samples of $n$
random PAs, drawn from a uniform distribution, and calculate the fraction of
samples with the HR statistic $T$ at least as extreme as the
observations. Note smaller $T$ statistics are more significant (opposite to
KS and Kuiper's tests).

Our implementation of the HR test does not currently incorporate
weighting. The HR test requires circular data so we apply it to 2-axial and
4-axial PAs after transformations $\Theta_{2ax} = 2\theta$ and $\Theta_{4ax}
= 4\theta_{4ax}$ respectively (see Appendix~\ref{appAxial}).

\subsubsection{$\chi^2$ test}
\label{subsubChi2}

The $\chi^2$ test will have lower discriminatory power than tests applied to
continuous data, such as Kuiper's and HR tests. Unlike those tests, it does
{\it not} account for the `wrap-around' nature of circular/axial data. It is
included predominantly due to its ease of computation and interpretation.

For a histogram comprising $m$ bins, the $\chi^2$ statistic is

\begin{ceqn}
\begin{equation} \label{eqChi2}
    \chi^2 = \sum_{i=1}^{m} \frac{(O_i - E_i)^2}{E_i}\ ,
\end{equation}
\end{ceqn}

\noindent where $O_i$ is the observed frequency and $E_i$ is the expected
frequency (in this case uniform) per bin $i$. To incorporate weighting we
compute the frequencies of a weighted histogram. Either $O_i$ and $E_i$ must
both be normalized, or, more simply and equivalently, the weighted
frequencies $O_{i,w}$ must be scaled such that

\begin{ceqn}
\begin{equation} \label{eqChi2Weights}
    \sum_{i=1}^{m} O_{i,w} = n\ ,
\end{equation}
\end{ceqn}

\noindent where $n$ is the total number of measurements in all $m$ bins.

We implement the $\chi^2$ test using \texttt{scipy\footnote{SciPy community
    project \citep{Virtanen2020}}.stats.chisquare}, which evaluates the test
statistic plus a $p$-value.

We apply the $\chi^2$ test to both unweighted and weighted PA histograms, of
both 2-axial and 4-axial PAs. For weighted histograms the observed frequency
$O_{i,w}$ is scaled (Eq.~\ref{eqChi2Weights}) and the expected frequency
$E_i$ is uniform and unweighted. In all cases the bin width is chosen to
ensure $E_i > 5$.

\subsection{Bimodality tests}
\label{subBimodalityMethods}

As for uniformity tests, we perform bimodality tests on PAs after parallel
transport to the centre of the A1 region \citep{Hutsemekers1998}.

To examine bimodality we considered Hartigans' dip statistic
\citep[HDS,][]{Hartigan1985}, the bimodality coefficient
\citep[BC,][]{SAS2004} and Akaike's information criterion difference
\citep[AIC\textsubscript{\emph{diff}},][]{Akaike1974}. \citet{Freeman2013}
compare these measures, and report that HDS has the highest sensitivity,
followed by BC, and that both methods are generally convergent. They found
that AIC\textsubscript{\emph{diff}} behaves quite differently, and
erroneously identifies bimodality in their simulations and experimental data.

We note the bimodality coefficient is unsuitable for hypothesis significance
testing and has an undesirable sensitivity to skew. We also found it
inconsistent with different bin sizes and concluded it was too capricious for
us to draw any conclusions from its results.

We therefore test the PA distribution for bimodality using Hartigans' dip
statistic.

\subsubsection{Hartigans' dip statistic}
\label{subsubHDS}

Hartigans' dip statistic \citep{Hartigan1985} is a non-parametric test of the
unimodality of continuous data. A distribution is categorised as unimodal if
its cumulative distribution function is convex up to its maximum gradient
(which corresponds to the peak in the distribution) and concave afterwards,
i.e.\ with a single inflection point. HDS quantifies how far the CDF departs
from unimodality, and indicates the location(s) of any departure, i.e.\ the
peak(s) in a bimodal (multimodal) distribution. This is well explained and
illustrated by \citet{Maurus2016}.

We evaluate HDS using Benjamin Doran's Python port of
\texttt{unidip.UniDip}\footnote{\url{https://github.com/BenjaminDoran/unidip}},
which follows \citet{Maurus2016}. The sensitivity of this test is controlled
by the parameter $\alpha$; we use $\alpha = 0.03$ to isolate peaks with at
least $97\%$ signal-to-noise confidence. This implementation does not
currently accommodate weighted data.

We evaluate Hartigans' dip statistic for continuous unweighted 2-axial
PAs. We do not apply it to 4-axial PAs, since this conversion yields unimodal
data, for which HDS is not meaningful.

\subsection{Correlation tests}
\label{subCorrelationMethods}

Uniformity and bimodality tests do not quantify the degree of alignment in
the data; for this we need specific statistical methods appropriate to axial
data on the celestial sphere. We considered two tests, the S~test and the
Z~test, that have been widely used to analyse quasar-polarisation alignments
(e.g.\ \citet{Hutsemekers1998}, \citet{Jain2004}, \citet{Pelgrims2015}), but
are appropriate to analyse the alignment of any vectors on the celestial
sphere (e.g.\ \cite{Contigiani2017}).

Our sample of 71 LQG PAs is relatively small. \citet{Hutsemekers2014} report
the Z~test is better suited to small samples than the S~test, because the
latter uses a measure of angle dispersion which suffers reduced power with
small samples. However, if PA alignment is `global' (i.e.\ correlations are
present throughout the survey area), then PAs will not be correlated to
positions and the power of the Z~test will reduce dramatically (Pelgrims
2019, private communication).

We find that LQG PAs are not correlated with position, and that the S~test is
more appropriate than the Z~test to quantify their alignment.

\subsubsection{S test}
\label{subsubSTest}

The S~test was developed by \citet{Hutsemekers1998} and analyses the
dispersion of vectors with respect to their nearest $n_v$ neighbours
(identified as explained in Appendix~\ref{subSTestNeighbours}). For each
vector~$i$ a measure of the dispersion $d_i$ is calculated as

\begin{ceqn}
\begin{equation} \label{eqDispOrig}
    d_i(\theta) = 90 - \frac{1}{n_v} \sum_{k=1}^{n_v}|90 - |\theta_k - \theta||\ ,
\end{equation}
\end{ceqn}

\noindent where $\theta_k$ are the 2-axial PAs [$0^{\circ}$, $180^{\circ}$)
  of the neighbouring $n_v$ vectors, including central vector $i$. The value
  of $\theta$ that minimises the function $d_i(\theta)$ is a measure of the
  average PA at the location of $i$. Use of absolute values accounts for the
  axial nature of the data \citep{Fisher1993}.

For vector $i$ the mean dispersion $D_i$ of its $n_v$ nearest neighbours is
calculated to be the minimum value of $d_i(\theta)$, which will be small for
coherently aligned vectors. The measure of alignment within the whole sample
of $n$ vectors is given by the S~test statistic

\begin{ceqn}
\begin{equation} \label{eqSD}
    S_D = \frac{1}{n} \sum_{i=1}^{n}D_i\ ,
\end{equation}
\end{ceqn}

\noindent with one free parameter $n_v$. If the vectors are aligned, the
value of $S_D$ will be smaller than if they are uniformly distributed. So,
the significance level for this version of the S~test is evaluated as the
probability that a random numerical simulation has a lower $S_D$ than that
observed \citep{Cabanac2005}.

\citet{Jain2004} introduce a coordinate invariant version of the S~test,
similar to the original except that, instead of the dispersion measure in
Eq.~\ref{eqDispOrig}, they use

\begin{ceqn}
\begin{equation} \label{eqDispJain}
    d_i(\theta) = \frac{1}{n_v} \sum_{k=1}^{n_v}\cos[2\theta - 2(\theta_k + \Delta_{k \rightarrow i})]\ ,
\end{equation}
\end{ceqn}

\noindent where $\Delta_{k \rightarrow i}$ is the angle by which the PA
$\theta_k$ changes during parallel transport from position $k$ to position
$i$. Here, the factor two accounts for the axial nature of the data. The
measure of dispersion is given by the \emph{maximum} value of
Eq.~\ref{eqDispJain} (as opposed to the minimum value of
Eq.~\ref{eqDispOrig}). The S statistic is calculated as previously
(Eq.~\ref{eqSD}). \citet{Pelgrims2016thesis} notes that the same value of
$\theta$ that maximises Eq.~\ref{eqDispJain} at the same time minimises
Eq.~\ref{eqDispOrig}, so the two versions are fully equivalent.

\citet{Jain2004} show the maximisation of $d_i(\theta)$ is calculated
analytically as

\begin{ceqn}
\begin{equation} \label{eqJainDiMax}
    d_i\Bigr\rvert_{max} = \frac{1}{n_v}\left[\left(\sum_{k=1}^{n_v}\cos\theta_k^{'}\right)^2 + \left(\sum_{k=1}^{n_v}\sin\theta_k^{'}\right)^2\right]^{1/2}\ ,
\end{equation}
\end{ceqn}

\noindent where $\theta_k^{'} = 2(\theta_k + \Delta_{k \rightarrow i})$ is
the circular version of $\theta_k$ after parallel transport to position
$i$. We can similarly apply this to the 4-axial version of $\theta_k$ by
using a factor of 4. This calculation is straightforward to code and avoids
the time-consuming trials of the original version \citep{Hutsemekers1998}. A
large value of $d_i\rvert_{max}$ indicates small dispersion, so a large value
of $S_D$ indicates strong alignment.

The significance level of the S~test for alignment (orthogonality) is the
probability that a random numerical simulation has a higher (lower) $S_D$
than that observed. See section~\ref{subSTestRandoms} for an explanation of
this interpretation, and details of how we estimate significance level using
numerical simulations.

We evaluate the S~test for unweighted 2-axial and 4-axial PAs. For 2-axial we
analyse PAs of the form $\Theta_{2ax} = 2\theta$, and for 4-axial we analyse
PAs of the form $\Theta_{4ax} = 4\theta_{4ax}$ (see
Appendix~\ref{appAxial}). In both cases we apply parallel transport
corrections before transforming the angles.

\section{Results: LQG position angles}
\label{secPAResults}

\begin{figure} 
	\centering
    \includegraphics[width=\columnwidth]{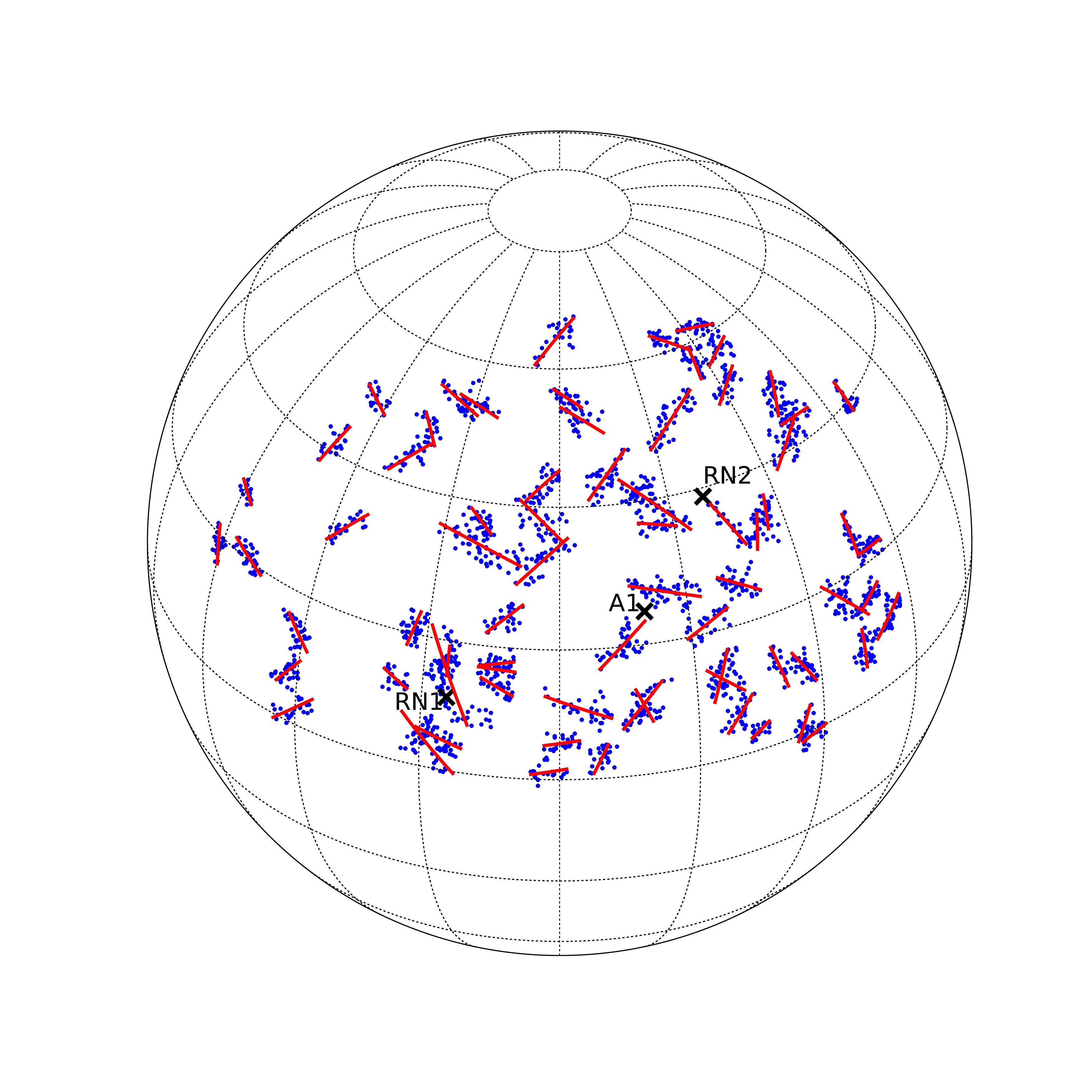}
    \caption[LQGs and their position angles on the celestial sphere]{LQG
      quasars (blue dots) and ODR axes (red lines) shown on the celestial
      sphere (east to the right). Also shown, the centres (black crosses) of
      A1, RN1, and RN2 regions \citep{Hutsemekers1998,Pelgrims2015}. A1 is
      the parallel transport destination for uniformity and bimodality
      tests. Projection is centred on $\alpha = 180^{\circ}$, $\delta =
      35^{\circ}$ (J2000), parallels and meridians are separated by
      $20^{\circ}$. RA increases to the right.}
    \label{figLQGsSphere}
\end{figure}

\begin{table*}
    \centering
    \caption[Large quasar group locations and position angles (example
      LQGs)]{Example large quasar groups (LQGs), where $m$ is the number of
      members, and $\bar{\alpha}$, $\bar{\delta}$, and $\bar{z}$ are the mean
      right ascension, declination, and redshift of the member quasars. The
      normalized goodness-of-fit weight $w$ (Eq.~\ref{eqWeight}) is scaled by
      $w_{71} = w \times 71$ for clarity, and to distinguish those LQGs
      weighted higher ($w_{71} > 1$) or lower ($w_{71} < 1$) than the mean
      $\bar{w}$. Position angle $\theta$ and half-width confidence interval
      $\gamma_h$ are shown for both the 2D and 3D approaches. The ratio of
      LQG ellipsoid axes lengths (from the 3D approach,
      Appendix~\ref{appLQGorientation}) is given by $a:b:c$. See
      Appendix~\ref{subappLQGorientation_tab} for the full sample of 71 LQGs.}
    \label{tabLQGDataPart}
    \begin{tabular}{crrrrrrrrc}
        \hline
        & \multicolumn{2}{r}{J2000 ($^\circ$)} &&& \multicolumn{2}{r}{2D PA ($^\circ$)} & \multicolumn{2}{r}{3D PA ($^\circ$)} & \\
        $m$ & $\bar{\alpha}$ & $\bar{\delta}$ & $\bar{z}$ & $w_{71}$ & $\theta$ & $\gamma_h$ & $\theta$ & $\gamma_h$ & $a:b:c$ \\
        \hline
        20 & 121.1 & 27.9 & 1.73 & 1.13 & 119.7 & 10.4 & 115.1 & 10.7 & 0.50:0.30:0.21 \\
        20 & 151.5 & 48.6 & 1.46 & 1.21 & 144.4 & 9.2 & 144.3 & 11.5 & 0.50:0.33:0.17 \\
        20 & 155.9 & 12.8 & 1.50 & 0.32 & 120.7 & 25.2 & 117.2 & 37.0 & 0.44:0.43:0.14 \\
        20 & 163.6 & 16.9 & 1.57 & 1.00 & 0.9 & 8.8 & 5.7 & 9.2 & 0.47:0.33:0.20 \\

        &&&&&\dots&&&& \\

        23 & 209.5 & 34.3 & 1.65 & 1.99 & 152.5 & 2.8 & 152.3 & 2.8 & 0.68:0.17:0.15 \\
        23 & 214.3 & 31.8 & 1.48 & 0.27 & 18.6 & 32.6 & 87.7 & 32.5 & 0.47:0.35:0.18 \\

        &&&&&\dots&&&& \\
        
        26 & 160.3 & 53.5 & 1.18 & 0.33 & 110.6 & 23.2 & 111.5 & 25.1 & 0.38:0.34:0.28 \\
        26 & 171.7 & 24.2 & 1.10 & 0.78 & 48.5 & 8.3 & 47.3 & 8.1 & 0.46:0.29:0.24 \\
        
        &&&&&\dots&&&& \\
        
        55 & 196.5 & 27.1 & 1.59 & 0.95 & 107.5 & 3.4 & 107.0 & 3.4 & 0.58:0.24:0.18 \\
        56 & 167.0 & 33.8 & 1.11 & 0.81 & 110.2 & 3.5 & 110.4 & 3.8 & 0.50:0.29:0.21 \\
        64 & 196.4 & 39.9 & 1.14 & 0.83 & 133.6 & 3.0 & 133.9 & 3.2 & 0.48:0.36:0.17 \\
        73 & 164.1 & 14.1 & 1.27 & 0.76 & 156.6 & 4.2 & 156.3 & 4.5 & 0.55:0.28:0.16 \\
        \hline
        \hline
    \end{tabular}
\end{table*}

\begin{figure*} 
    \centering
	\subfigure[Unweighted]{\includegraphics[width=0.7\columnwidth]{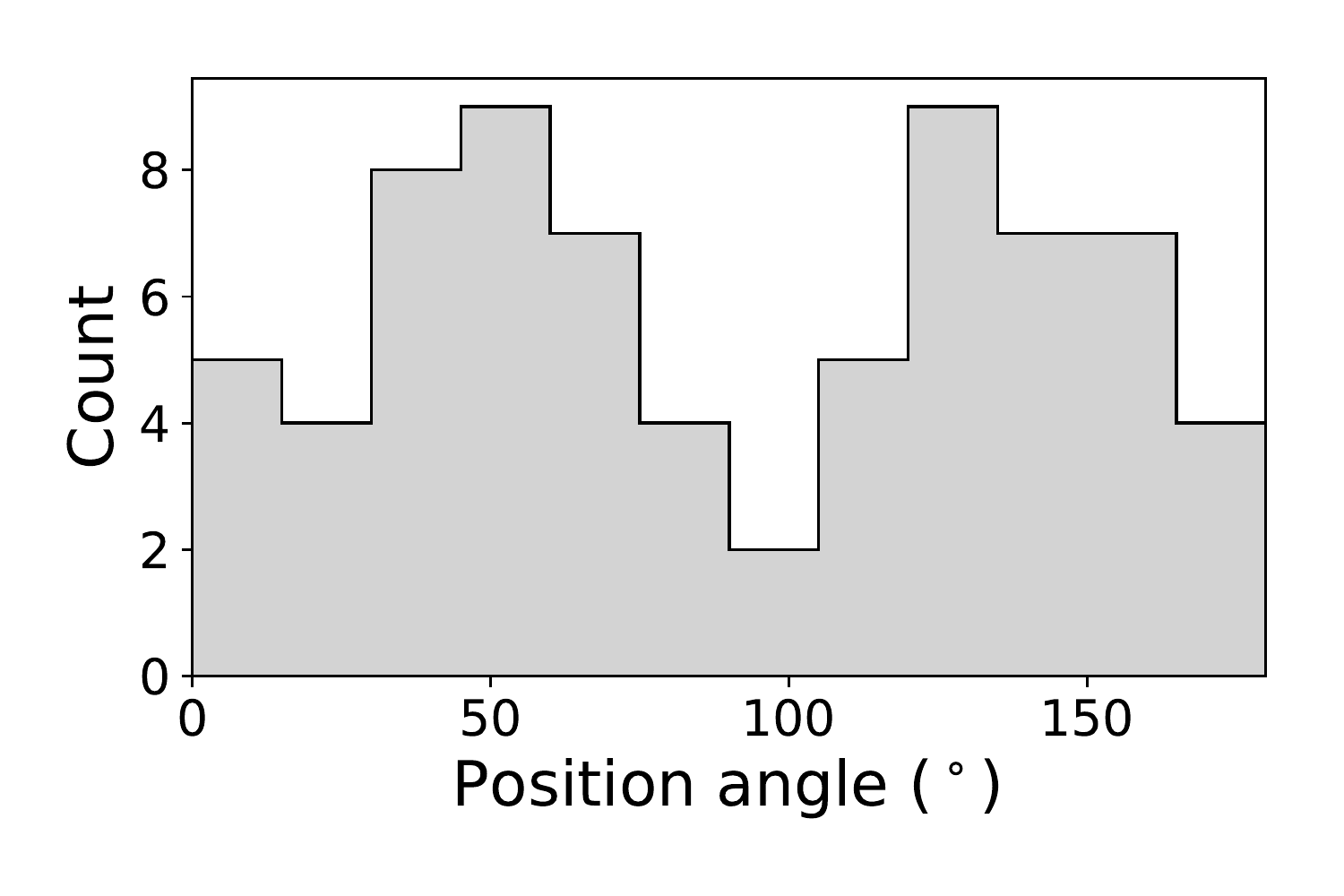} \label{subHistUnweighted}}
	\subfigure[Weighted]{\includegraphics[width=0.7\columnwidth]{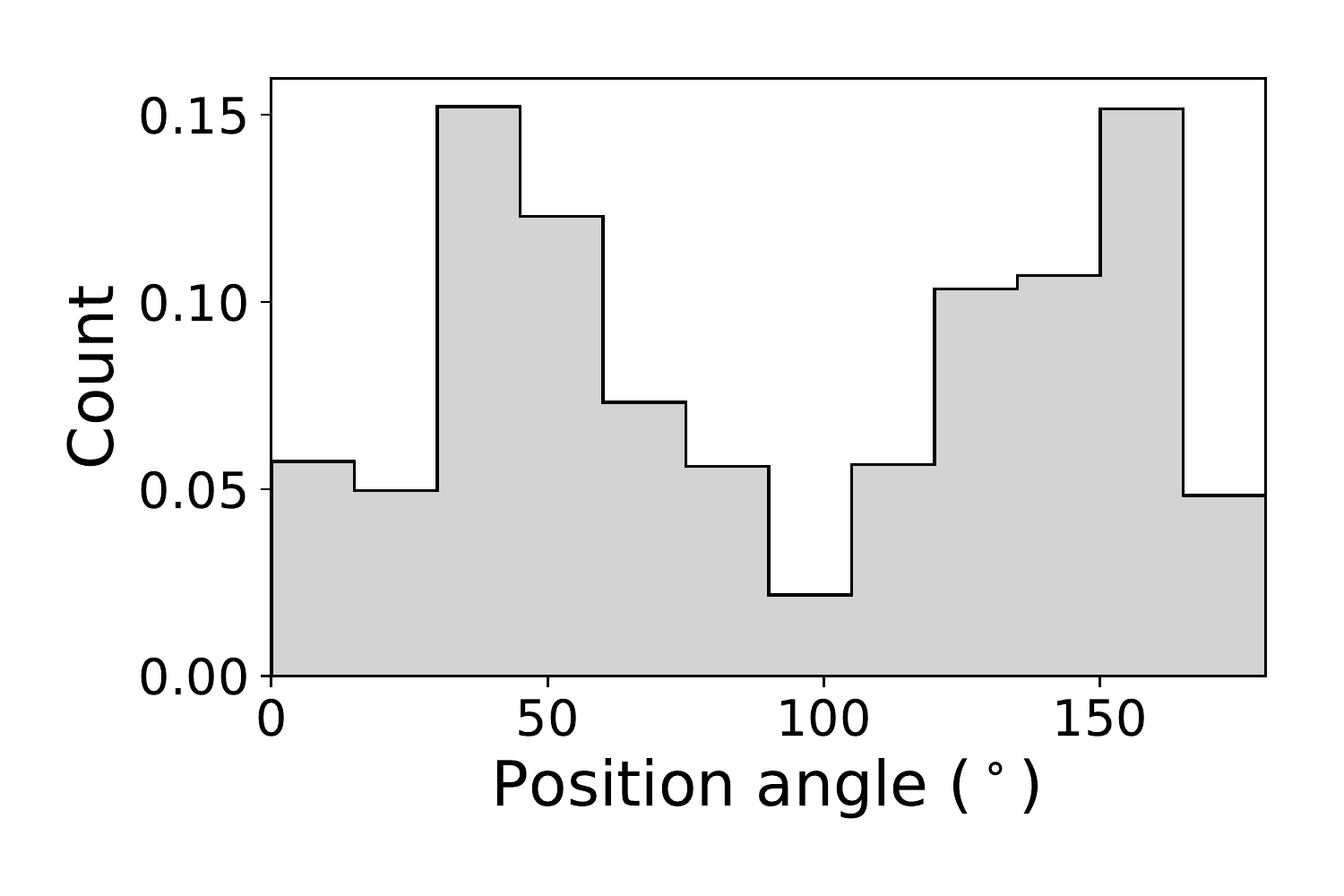} \label{subHistWeighted}}
	\caption[Position angles with and without weighting (histogram)]{LQG
          position angles, all parallel transported to and measured at the
          centre of the A1 region
          \citep{Hutsemekers1998}. \protect\subref{subHistUnweighted} is
          unweighted and \protect\subref{subHistWeighted} is ODR
          goodness-of-fit weighted, both with $15^{\circ}$ bins. The bimodal
          distribution is robust to whether or not ODR goodness-of-fit
          weighting is used.}
	\label{figPAHist}
\end{figure*}

\begin{figure*} 
    \centering
	\subfigure[Unweighted]{\includegraphics[width=0.7\columnwidth]{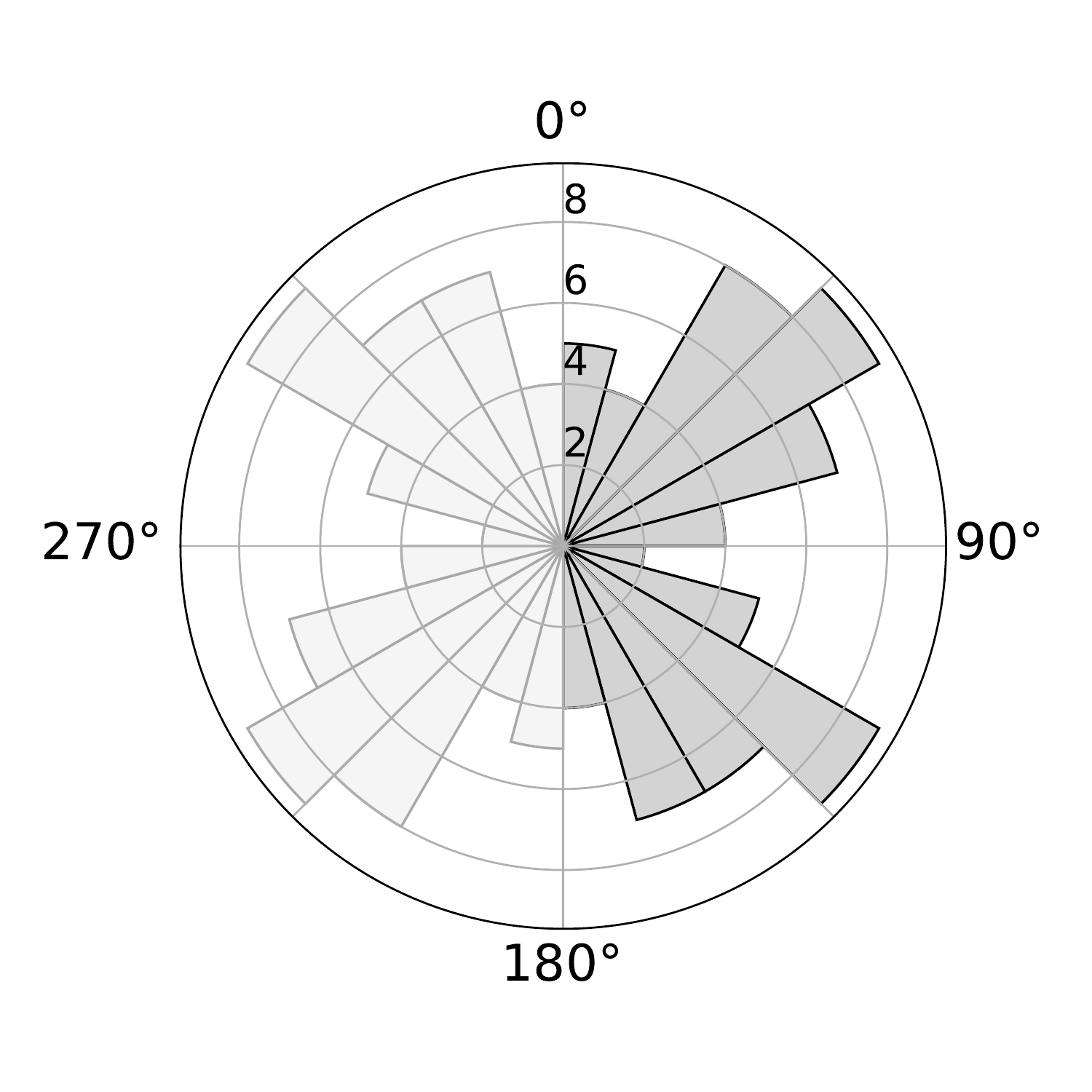} \label{subRoseUnweighted}}
	\subfigure[Weighted]{\includegraphics[width=0.7\columnwidth]{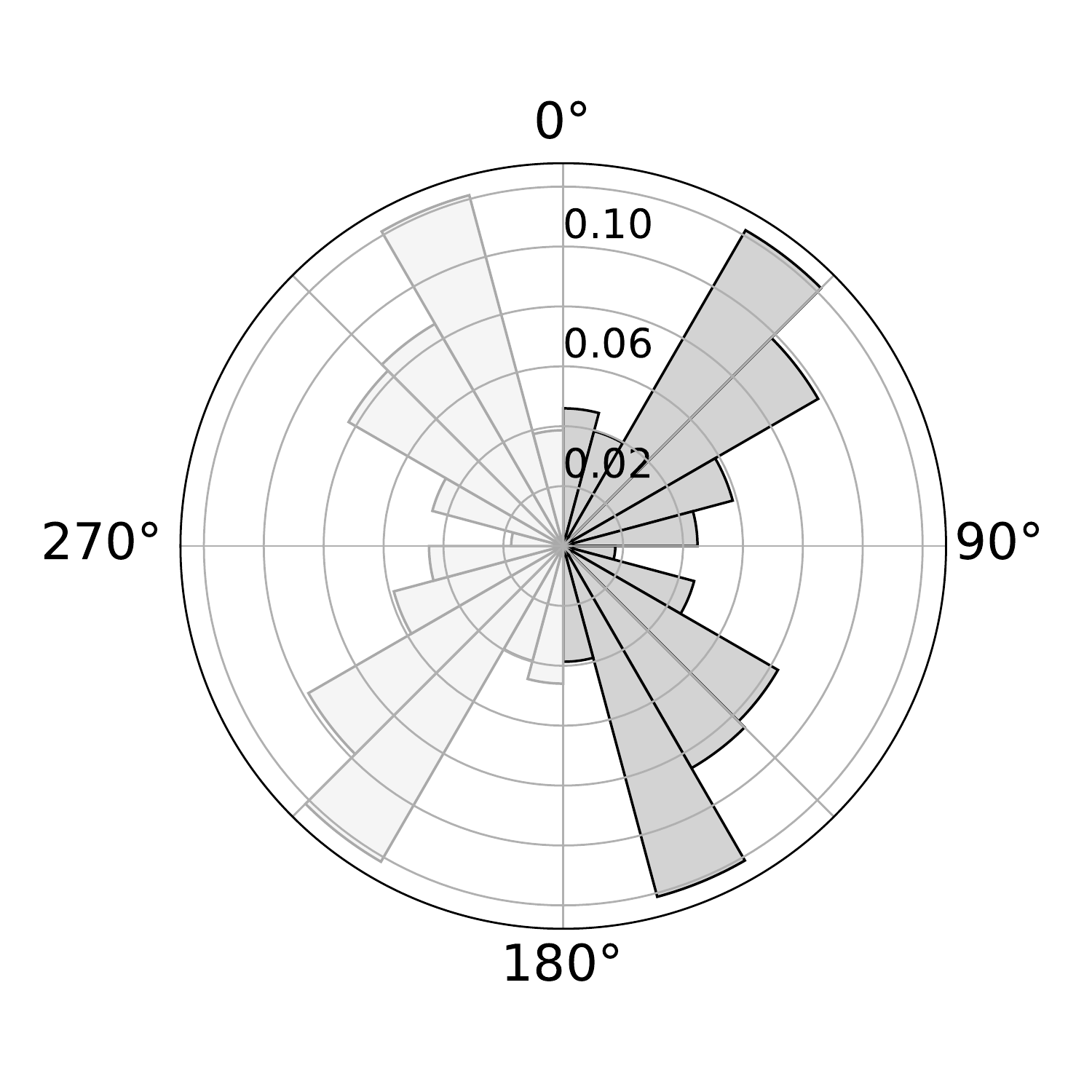} \label{subRoseWeighted}}
	\caption[Position angles with and without weighting (rose
          diagram)]{As Fig.~\ref{figPAHist} but represented as a rose
          diagram; again \protect\subref{subRoseUnweighted} is unweighted and
          \protect\subref{subRoseWeighted} is ODR goodness-of-fit weighted,
          both with $15^{\circ}$ bins. As is conventional for undirected
          axial data the [$0^{\circ}$, $180^{\circ}$) data are duplicated on
            the opposite side of the rose diagram [$180^{\circ}$,
              $360^{\circ}$). For clarity, we reiterate that the PA data are
              axial [$0^{\circ}$, $180^{\circ}$) and not circular
                [$0^{\circ}$, $360^{\circ}$); we do not have PAs in the range
                  [$180^{\circ}$, $360^{\circ}$) (the lighter shade).}
	\label{figPARose}
\end{figure*}

We identify 71 LQGs of $\ge 20$ quasars and detection significance $\ge
2.8\sigma$. The LQG positions on the celestial sphere, and their orientation
as determined by the two-dimensional method, are illustrated in
Fig.~\ref{figLQGsSphere}. By eye it appears that the orientations may be
somewhat preferentially aligned, but we caution that the orthographic
projection may be deceiving. LQG positions in three-dimensional proper
space, and their orientations, as determined by the three-dimensional method,
are illustrated in Appendix~\ref{appLQGs3DPerspectives}.

The results of both the 2D and 3D approaches are presented in
Tables~\ref{tabLQGDataPart} (example LQGs) and \ref{tabLQGDataFull} (full LQG
sample, Appendix~\ref{subappLQGorientation_tab}), and generally agree well. The PAs listed
in these tables are measured in situ at the location of each LQG, and will
have parallel transport corrections applied before statistical analysis
(section~\ref{secStatsResults}). For both approaches, bootstrap re-sampling
with replacement is used to estimate the uncertainty in the form of the
half-width confidence interval (HWCI, $\gamma_h$) of 10000 bootstraps.

Figs.~\ref{figPAHist} (histograms) and \ref{figPARose} (rose diagrams) show
LQG PAs, after parallel transport, both unweighted and weighted by orthogonal
distance regression goodness-of-fit (Eq.~\ref{eqWeight}). For axial data
[$0^{\circ}$, $180^{\circ}$), where $0^{\circ}$ and $180^{\circ}$ are
  equivalent, a conventional histogram (Fig.~\ref{figPAHist}) can be
  misleading, since it represents data that are close together
  (e.g.\ $1^{\circ}$ and $179^{\circ}$) at opposite extremes of the
  distribution. An alternative representation is the rose diagram
  (Fig.~\ref{figPARose}), where wedge length is proportional to the count and
  spanning angle denotes the bins.

In both Figs.~\ref{figPAHist} and \ref{figPARose} the data appear bimodal,
with peaks at $\theta\sim45^{\circ}$ and $\theta\sim135^{\circ}$ (in the
absence of goodness-of-fit weighting). The peaks are separated by
$\Delta\theta\sim90^{\circ}$, indicating that some LQGs may have PAs that are
preferentially parallel (i.e.\ aligned) while others are preferentially
orthogonal to one another (this is described as `anti-aligned' by
\citet{Hutsemekers2014} and \citet{Pelgrims2016thesis}).

\subsection{LQG PAs as a function of redshift}
\label{subLQGPARedshift}

\begin{figure} 
	\centering
    \includegraphics[width=0.8\columnwidth]{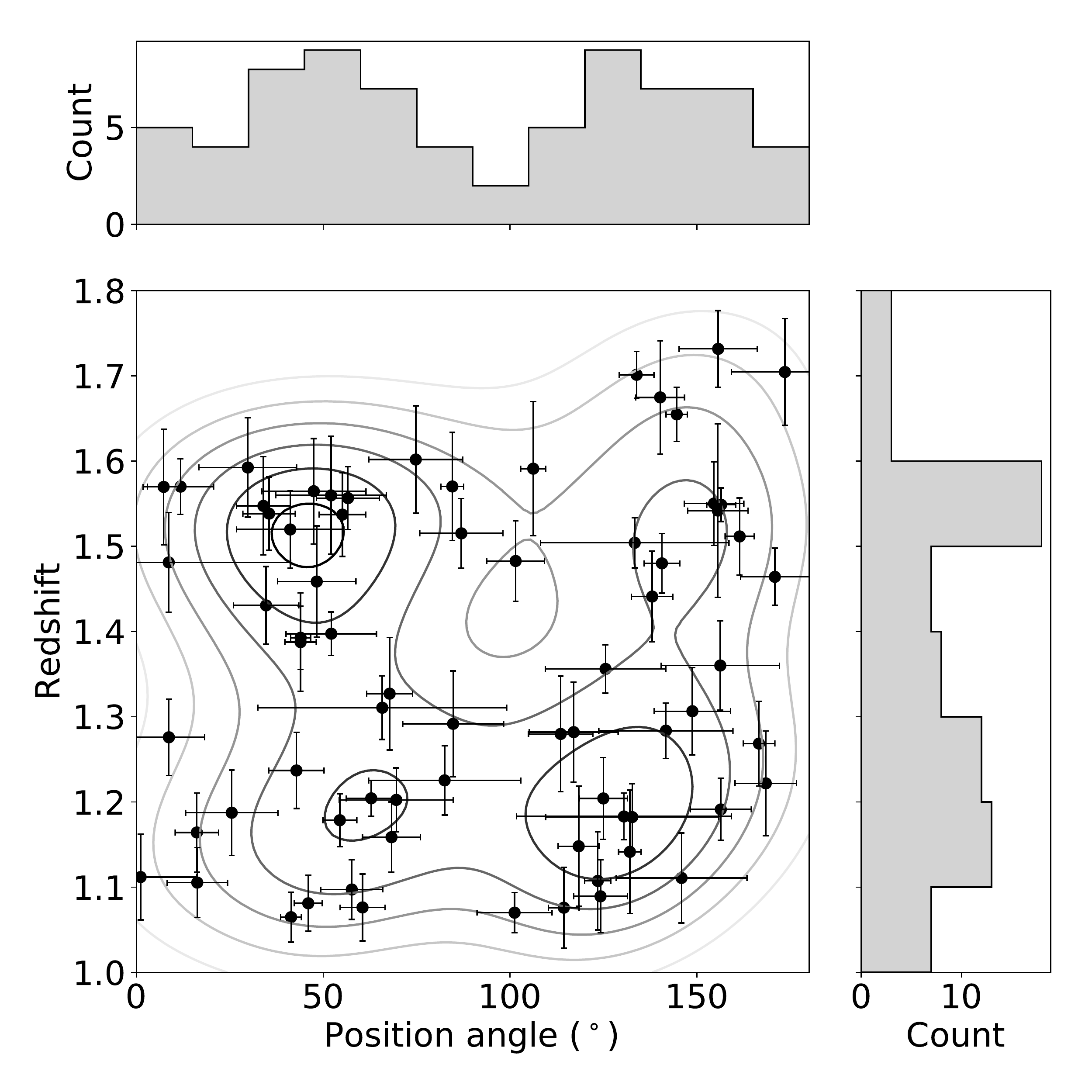}
    \caption[Marginal plot of the LQG PA $\times$ redshift plane]{Marginal
      plot of the LQG PA $\times$ redshift plane. Error bars are PA
      half-width confidence interval and redshift standard
      deviation. Marginal histograms show PA and redshift distributions with
      $\Delta \theta = 15^{\circ}$ and $\Delta z = 0.1$ bins. Contours are a
      Gaussian kernel density estimation of the scatter plot. Histograms and
      KDE are unweighted.}
    \label{figPAzMarginal}
\end{figure}

\begin{figure} 
	\centering
    \includegraphics[width=0.7\columnwidth]{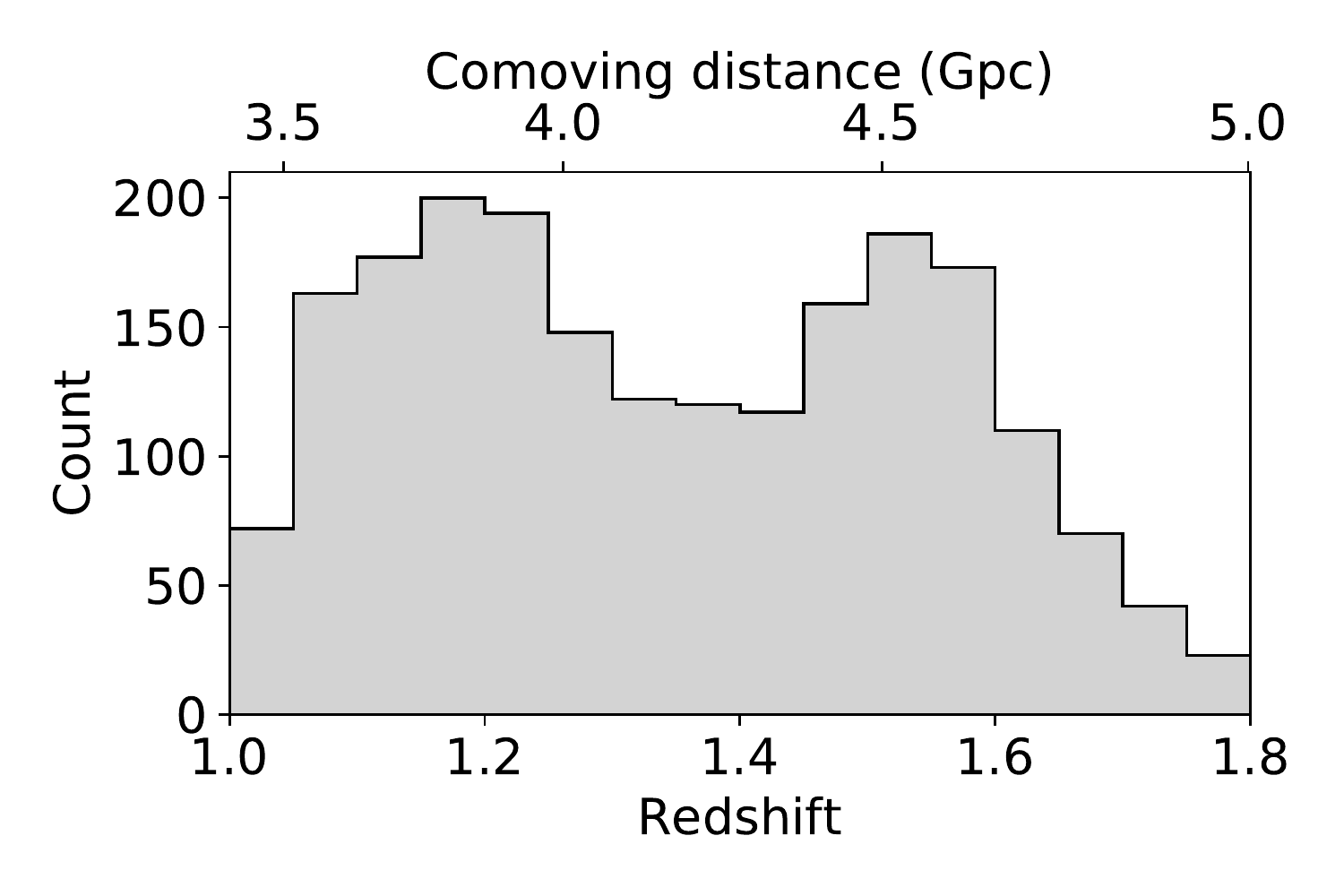}
    \caption[Redshift distribution of member quasars from all 71
      LQGs]{Redshift distribution of the 2076 member quasars comprising our
      sample of 71 LQGs, with $\Delta z = 0.05$ bins, unweighted. Top x-axis
      shows approximate proper radial distance. Distribution is bimodal,
      with modes (peaks) at $z \sim 1.15$ and $z \sim 1.55$.}
    \label{figQSOz}
\end{figure}

Fig.~\ref{figPAzMarginal} shows a marginal plot of the LQG PA $\times$
redshift plane; redshift here is LQG redshift, defined as the mean redshift
of its member quasars. From Fig.~\ref{subHistUnweighted}, we expect the PA
histogram (top margin) to be bimodal, as seen. The redshift histogram (right
margin) also exhibits some bimodality. To investigate the relationship
between these two variables, and whether there is any correlation between
their modes, we add kernel density estimation (KDE) contours to the scatter
plot. This shows hints of three or four modes, although the correlation is
weak. We note that the apparently stronger modes at $\theta \sim 45^{\circ}$
$\times$ $z \sim 1.5$ and $\theta \sim 135^{\circ}$ $\times$ $z \sim 1.2$
result from the points with the greatest uncertainty. Conversely, the weaker
modes at $\theta \sim 55^{\circ}$ $\times$ $z \sim 1.2$ and $\theta \sim
150^{\circ}$ $\times$ $z \sim 1.5$ result from the points with the smallest
uncertainty.

Due to their scale, LQGs extend considerably in the radial direction, and
some features may be lost when we analyse only their mean redshift. Our
sample of 71 LQGs collectively comprise 2076 member
quasars. Fig.~\ref{figQSOz} shows the redshift distribution of these. The
distribution appears bimodal with modes (peaks) at $z \sim 1.15$ and $z \sim
1.55$, similar to that for LQG redshifts (Fig.~\ref{figPAzMarginal}, right
margin).

\subsection{LQG PA weights}
\label{subLQGPAWeights}

\begin{figure} 
    \centering
	\subfigure[ODR GoF weighted]{\includegraphics[width=0.7\columnwidth]{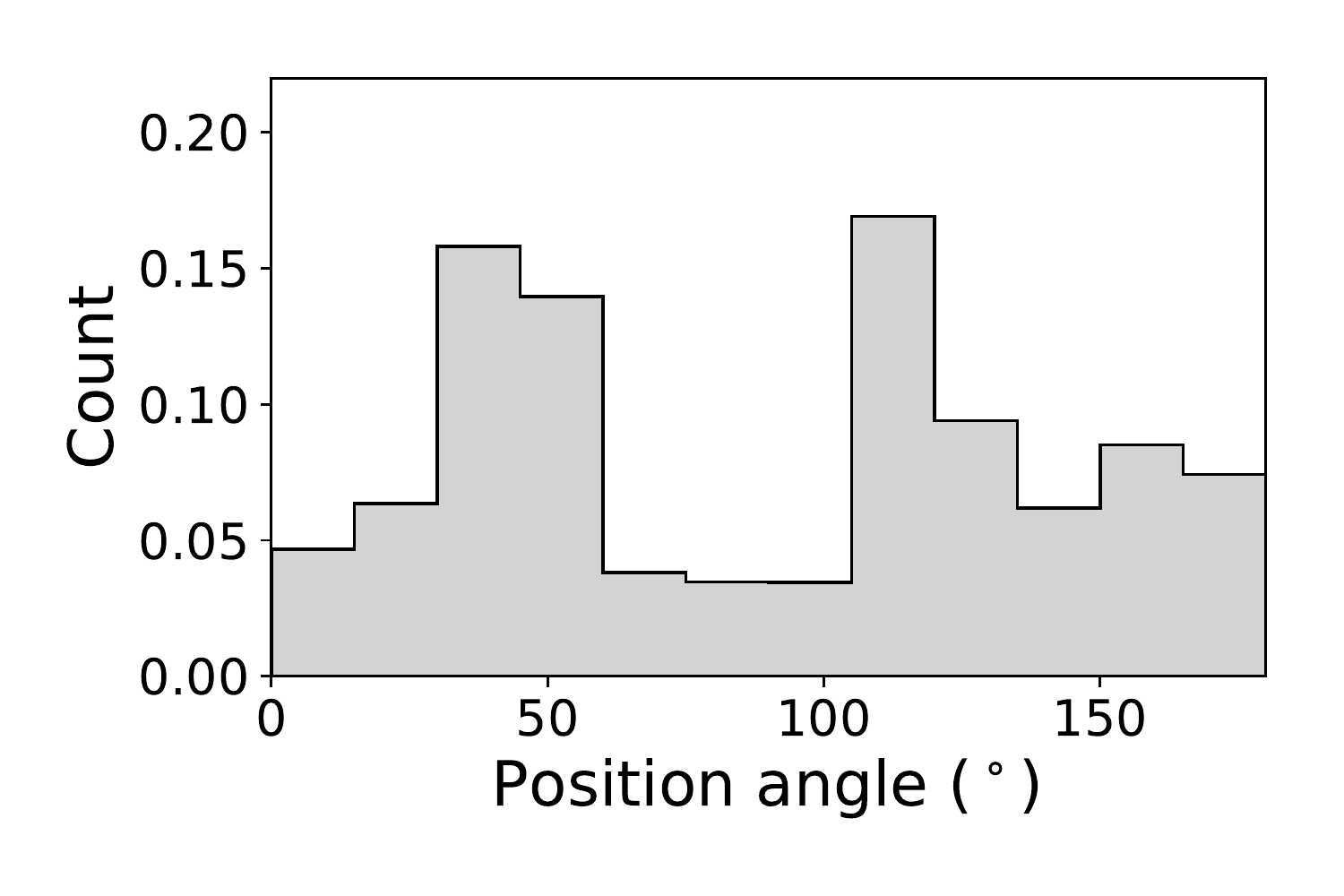} \label{subHistGoFWeighted}}\\
	\subfigure[HWCI weighted]{\includegraphics[width=0.7\columnwidth]{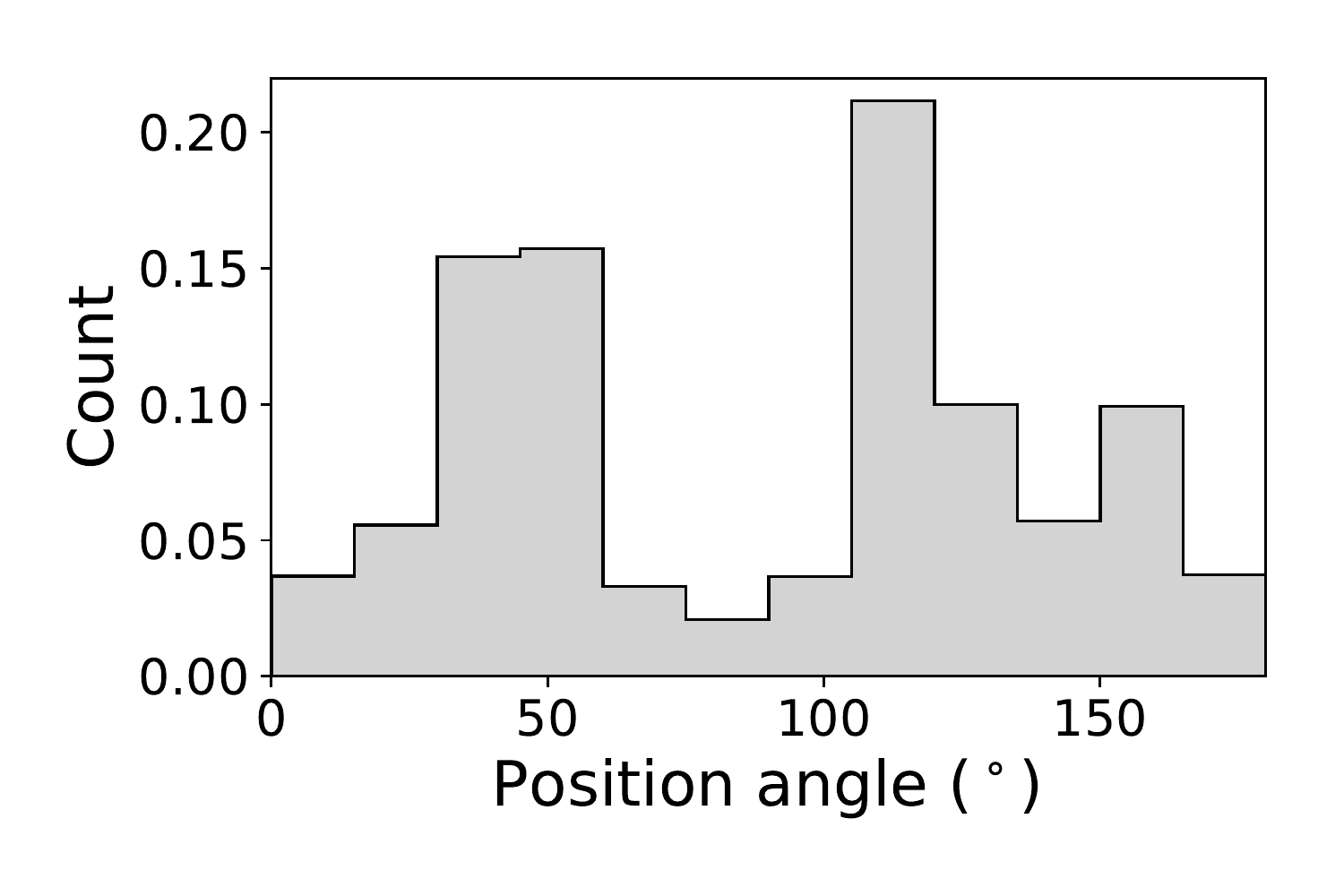} \label{subHistHWCIWeighted}}
	\caption[Position angles with alternative weighting
          (histogram)]{Position angles calculated by 2D approach, with no
          parallel transport, and $15^{\circ}$
          bins. \protect\subref{subHistGoFWeighted} is goodness-of-fit
          weighted and \protect\subref{subHistHWCIWeighted} is half-width
          confidence interval weighted. The bimodal distribution is robust to
          which method of weighting is used.}
	\label{figWeightedHist}
\end{figure}

The two-dimensional approach of determining position angles uses orthogonal
distance regression (ODR) of tangent plane projected quasars. We evaluate the
ODR goodness-of-fit (Eq.~\ref{eqWeight}), which may be used to weight the PAs
used for some of the statistical analysis (section~\ref{secStatsResults}). An
alternative empirical weighting scheme could use measurement uncertainties,
or half-width confidence intervals (HWCIs,
Appendix~\ref{appPAuncertainties}), estimated from 10000 bootstraps.

The PA distribution is robust between these alternative weighting
schemes. Fig.~\ref{figWeightedHist} shows PAs, determined by the 2D approach,
with no parallel transport, and weighted by
\protect\subref{subHistGoFWeighted} goodness-of-fit weights $w = \ell /
\sigma^2$, and \protect\subref{subHistHWCIWeighted} HWCI weights $w = 1 /
\gamma_h$. The bimodal distribution is robust to which method of weighting is
used. We continue to use the former in this work as it is more physically
motivated and slightly more conservative.

\section{Results: statistical analysis}
\label{secStatsResults}

The position angle distribution of our sample of 71 LQGs appears bimodal,
with modes at $\bar{\theta}\sim52\pm2^{\circ}, 137\pm3^{\circ}$ (with
goodness-of-fit weighting) after parallel transport to the centre of the A1
region \citep{Hutsemekers1998}. The median location of the peaks after
parallel transport to all 71 LQG locations is
$\bar{\theta}\sim45\pm2^{\circ}, 136\pm2^{\circ}$. The peaks are separated by
$\Delta\theta\sim90^{\circ}$, indicating that some LQGs have PAs that are
preferentially aligned with each other, while others are preferentially
orthogonal. We apply the statistical methods of
section~\ref{secLQGStatsMethods} to analyse the uniformity, bimodality, and
correlation of these PAs.

\subsection{Uniformity: LQG PAs are unlikely to be uniform}
\label{subsecUniformityResults}

\begin{table} 
    \centering
    \caption[Results of uniformity tests ($\chi^2$, Kuiper's,
      Hermans-Rasson)]{Results ($p$-values) of uniformity tests. For 2-axial
      (4-axial) PAs the $\chi^2$ test is evaluated using $20^{\circ}$
      ($10^{\circ}$) bins. The $\chi^2$ and Kuiper's tests are computed both
      with and without goodness-of-fit weighting. For 2-axial PAs, only the
      $\chi^2$ test shows evidence for non-uniformity (of weighted PAs). For
      4-axial PAs, all tests show evidence for non-uniformity, mostly
      marginal, with Kuiper's being the most significant.}
    \label{tabUniformityStats}
    \begin{tabular}{l c c c c}
        \hline
        & \multicolumn{4}{c}{$p$-value} \\
        & \multicolumn{2}{c}{2-axial} & \multicolumn{2}{c}{4-axial} \\
        test & unweighted & weighted & unweighted & weighted \\
        \hline
        $\chi^2$ & 0.16 & 0.01 & 0.07 & 0.02 \\
        Kuiper's & 0.62 & 0.59 & 0.009 & 0.008 \\
        HR & 0.07 & - & 0.04 & - \\
        \hline
    \end{tabular}
\end{table}

The results from applying the uniformity tests are listed in
Table~\ref{tabUniformityStats}. The $\chi^2$ test does not show evidence of
non-uniformity for 2-axial PAs without weighting ($16\%$ significance level),
but does show some evidence of non-uniformity with weighting ($1\%$
significance level). For 4-axial PAs it shows marginal evidence of
non-uniformity both with and without weighting ($2\%$ and $7\%$
respectively).

Kuiper's test does not show evidence of non-uniformity for 2-axial PAs, with
or without weighting. However, the PA distribution is bimodal, which
dramatically reduces the test's discriminatory power, so the absence of a
signal is unsurprising. For 4-axial PAs Kuiper's test indicates a rejection
of the null hypothesis of uniformity at the $0.8\%$ and $0.9\%$ significance
level (weighted and unweighted), i.e.\ $\sim 2.4\sigma$.

Finally, the Hermans-Rasson test shows marginal evidence of non-uniformity
for 2-axial and 4-axial PAs ($7\%$ and $4\%$ significance levels,
respectively), both without weighting.

Based on the results from these three uniformity tests we cannot confidently
reject the null hypothesis that the observed PAs are drawn from a uniform
distribution. The most appropriate test for the axial and bimodal nature of
the 2-axial PA data is the HR test, which shows marginal evidence of
non-uniformity.

However, if we \emph{a priori} expect \emph{f}-fold symmetry (specifically
2-fold, in case of a bimodal distribution) then the conversion of PAs to
4-axial becomes physically well motivated as well as statistically
legitimate. In this case, based on the 4-axial results of Kuiper's test, we
could confidently reject the null hypothesis and conclude that the PAs are
non-uniform.

\subsubsection{Uniformity of mock LQG catalogues}

\begin{table}
    \centering
    \caption[Results of uniformity tests applied to mock LQGs]{Results
      ($p$-values) of uniformity tests applied to mock LQGs. For 2-axial
      (4-axial) PAs the $\chi^2$ test is evaluated using $15^{\circ}$
      ($7.5^{\circ}$) bins, except for individual mocks where it is evaluated
      using $30^{\circ}$ ($15^{\circ}$) bins. For multiple samples the number
      of LQGs and $p$-values are means plus the standard error on the
      mean. All PAs are unweighted and are parallel transported to, and
      measured at, the centre of the Al region \citep{Hutsemekers1998}. Most
      samples and tests show no evidence for non-uniformity.}
    \label{tabMockChi2}
    \begin{tabular}{c l l l l}
         \hline
         & No. of & & \multicolumn{2}{c}{$p$-value} \\
         sample(s) & LQGs & test & \multicolumn{1}{c}{2-axial} & \multicolumn{1}{c}{4-axial} \\
         \hline
         110 & \multirow{3}{*}{$30\pm 0.4$} & $\chi^2$ & $0.53 \pm 0.03$ & $0.51 \pm 0.03$ \\
         individual & & Kuiper's & $0.53 \pm 0.03$ & $0.50 \pm 0.03$ \\
         mocks & & HR & $0.48 \pm 0.03$ & $0.50 \pm 0.03$ \\
         \arrayrulecolor{lightgray}\hline
         10 stacks & \multirow{3}{*}{$330\pm 3$} & $\chi^2$ & $0.47 \pm 0.10$ & $0.41 \pm 0.08$ \\
         of 11 & & Kuiper's & $0.56 \pm 0.10$ & $0.44 \pm 0.09$ \\
         sub-volumes & & HR & $0.46 \pm 0.11$ & $0.62 \pm 0.09$ \\
         \arrayrulecolor{lightgray}\hline
         1 stack & \multirow{3}{*}{3,296} & $\chi^2$ & 0.20 & 0.31 \\
         of & & Kuiper's & 0.27 & 0.03 \\
         110 mocks & & HR & 0.16 & 0.15 \\
         \arrayrulecolor{black}\hline
    \end{tabular}
\end{table}

The results from applying uniformity tests to the mock LQGs of
section~\ref{subMockLQGs} are listed in Table~\ref{tabMockChi2}. The $\chi^2$
and Hermans-Rasson tests show no evidence for non-uniformity of the mock
LQGs. This result is consistent between 2 and 4-axial PAs, individual mocks,
stacks of sub-volumes, and the stack of all 3,296 mock LQGs.

Kuiper's test also shows no evidence for non-uniformity, except for the stack
of all mock LQGs evaluated as 4-axial data (and then only marginally). This
could be an artefact of the realizations not being truly independent, but if
so it is unclear why this would reveal itself only in one of the six tests on
this sample. Our concerns about the independence of this particular stack,
the otherwise highly consistent results, and our caution about interpreting
results manifest only in 4-axial data, led to this anomalous result being
discredited. We therefore conclude that all three tests indicate statistical
uniformity of the mock LQGs.

\subsection{Bimodality: LQG PA distribution is bimodal}
\label{subsecBimodalityResults}

\begin{figure} 
    \centering
	\includegraphics[width=0.7\columnwidth]{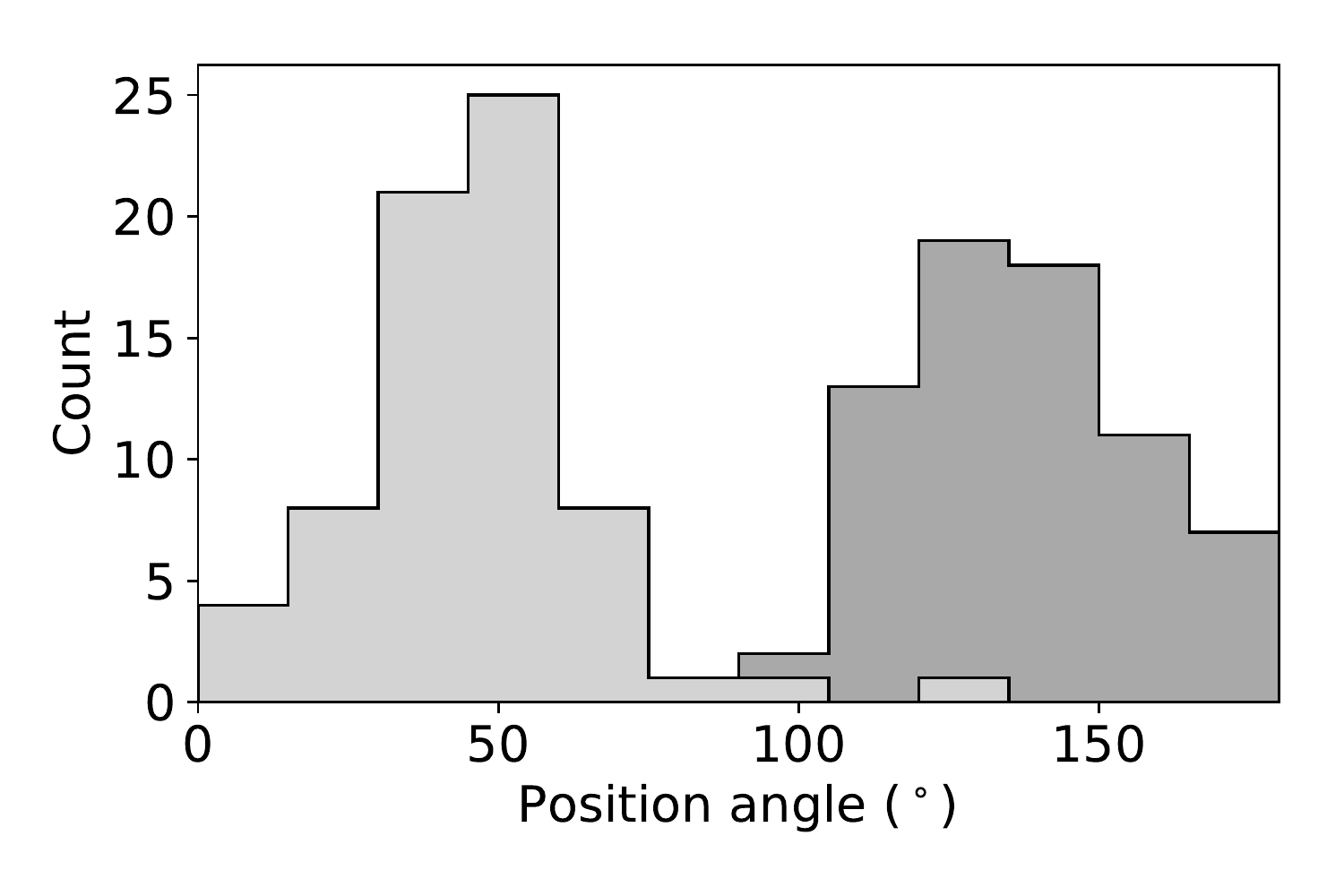}
	\caption[Location of double Gaussian peaks from
          Fig.~\ref{figGaussians}]{Location of the centres of the double
          Gaussian peaks in Fig.~\ref{figGaussians}; $15^{\circ}$ bins. The
          locations of the two peaks are consistent, regardless of parallel
          transport destination. The mean locations are $\sim47\pm 2^{\circ}$
          and $\sim138\pm 2^{\circ}$, consistent (within
          $\lesssim10^{\circ}$) with the HDS result.}
	\label{figPeaks}
\end{figure}

We calculate Hartigans' dip statistic (HDS) for continuous unweighted PA data
and recover two peaks between $\sim36^{\circ} - 83^{\circ}$ and
$\sim 114^{\circ} - 156^{\circ}$, with 97\% confidence. The unweighted
(weighted) means of PAs in these two ranges are $\sim 54\pm2^{\circ}$
($\sim 52\pm2^{\circ}$) and $\sim 136\pm3^{\circ}$ ($\sim 137\pm3^{\circ}$),
consistent with the peaks we see in the distribution of categorical PA data
(e.g.\ Fig.~\ref{figPAHist}). The bimodality is therefore unlikely to be an
artefact of binning.

We apply HDS after all PAs are parallel transported to the centre of the A1
region. Recalling section~\ref{subParallelTransport} \citep[see
  also][]{Jain2004}, the process of parallel transport rotates PAs; the
amount of rotation depending on the path taken (direction and
distance). Therefore it is reasonable to check whether parallel transporting
to a different location would affect the position of the PA peaks.

We parallel transport all 71 PAs to the location of each of the 71 LQGs, and
at each one fit a double Gaussian to the unweighted histogram of these
PAs. The height, width, and location of each Gaussian are fitted using
Python's \texttt{scipy.optimize.leastsq}.

The location of the centres of the double Gaussian peaks, at each of the 71
LQG locations, are shown in Fig.~\ref{figPeaks}. The median (mean) location
of the first peak is $\sim45\pm2^{\circ}$ ($\sim47\pm2^{\circ}$), and of the
second is $\sim136\pm2^{\circ}$ ($\sim138\pm2^{\circ}$). These are consistent
(within the mean PA half-width confidence interval of $\sim10^{\circ}$) with
the peaks identified by HDS at the centre of the A1 region
($\sim54\pm2^{\circ}$ and $\sim136\pm3^{\circ}$ for unweighted PAs).

Individual PA distributions at each of the 71 LQG locations are shown in
Fig.~\ref{figGaussians} (Appendix~\ref{appGaussians}). At most ($\gtrsim
80\%$) the PAs show a similar bimodal distribution.

\subsection{Correlation: LQG PAs are aligned and orthogonal}
\label{subsecCorrelationResults}

\begin{figure} 
    \centering
    \includegraphics[width=0.98\columnwidth]{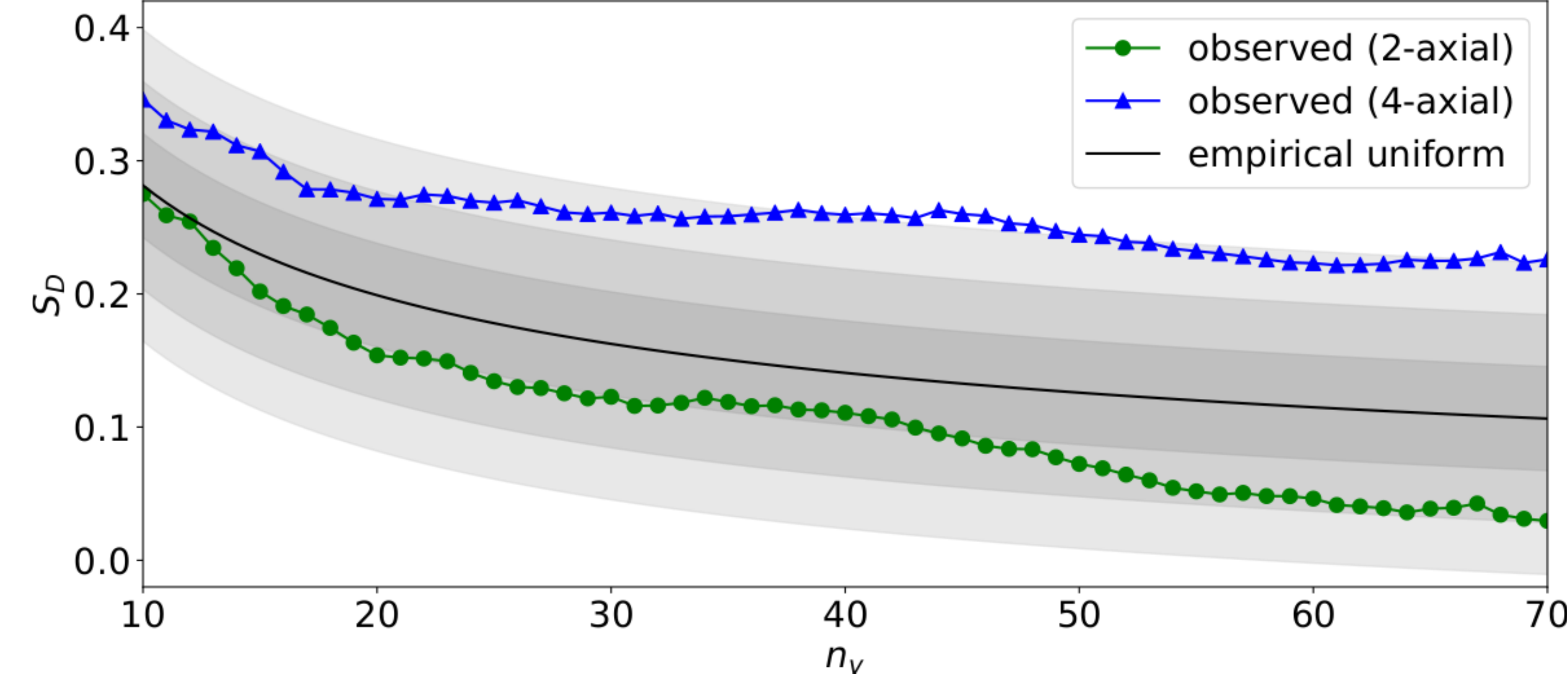}
    \caption[S~test statistic for 2-axial and 4-axial position angles]{The
      S~test statistic $S_D$ calculated for 2-axial (green circles) and
      4-axial (blue triangles) PAs as a function of nearest neighbours $n_v$
      determined in 3D. Also shown, empirical values estimated for a uniform
      distribution (black) and their approximate $\pm
      1\sigma,\ 2\sigma,\ 3\sigma$ confidence intervals (grey). 4-axial PAs
      show more alignment than uniform, while 2-axial show less.}
    \label{figSD}
\end{figure}

\begin{figure} 
    \centering
    \includegraphics[width=0.98\columnwidth]{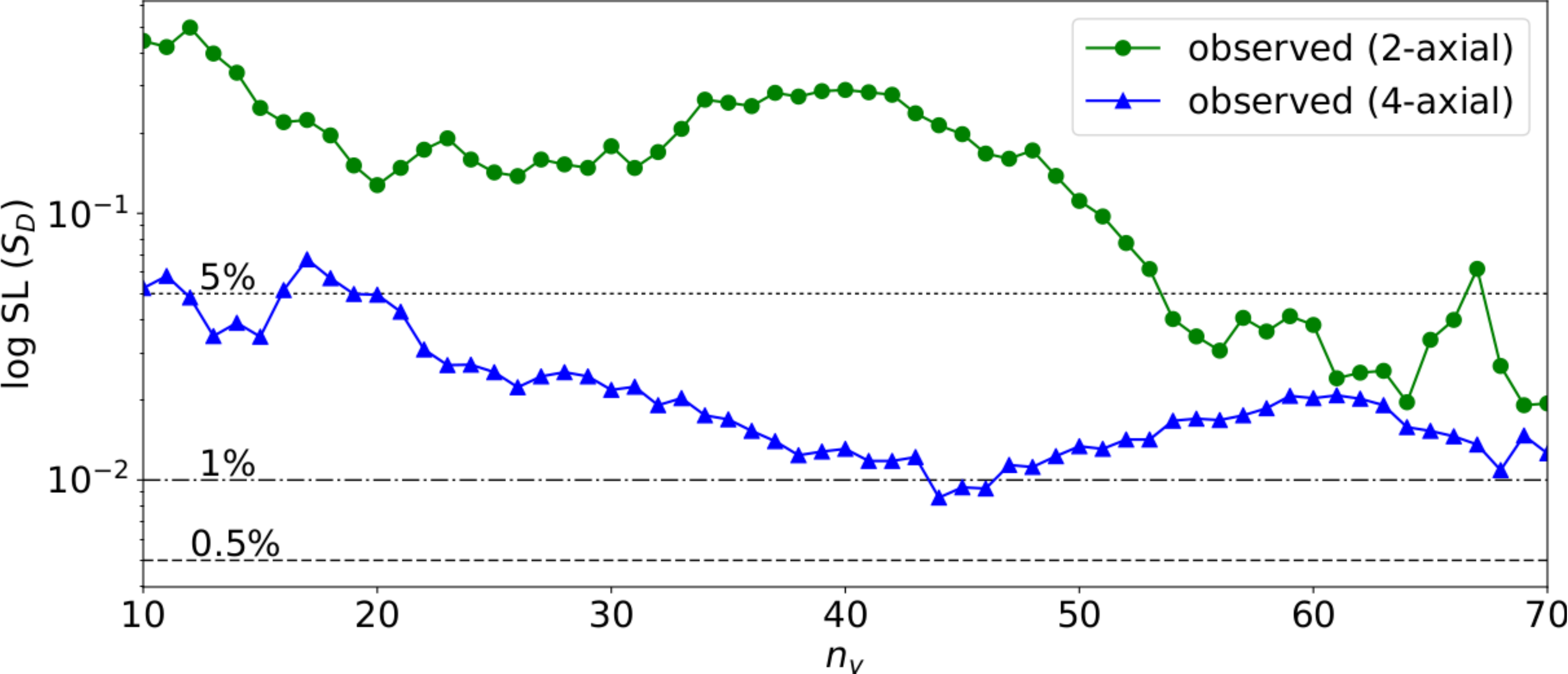}
    \caption[S~test significance level for 2-axial and 4-axial position
      angles]{The logarithmic significance level (SL) of the S~test
      calculated for 2-axial (green circles) and 4-axial (blue triangles) PAs
      as a function of nearest neighbours $n_v$ determined in 3D. The dotted,
      dash-dotted and dashed horizontal lines indicate SL = 0.05, 0.01 and
      0.005 respectively. 2-axial LQG PAs show correlation
      ($\mathrm{\overline{SL}}\sim 3.3\%$) above $n_v \ge 54$. 4-axial PAs
      show correlation ($\mathrm{\overline{SL}}\sim 1.5\%$) above $n_v \ge
      30$.}
    \label{figSL}
\end{figure}

We compute the S~statistic $S_D$ using the \citet{Jain2004} coordinate
invariant version of the S~test for samples of $n_v$ nearest neighbours,
where $10 \leq n_v \leq 70$. In Fig.~\ref{figSD} we show the values of $S_D$
calculated for observed LQG PAs, represented as both 2-axial (green circles)
and 4-axial (blue triangles) data, as a function of nearest neighbours
$n_v$. Nearest neighbours are determined in 3D, taking into account radial
distance. Results are entirely consistent with neighbours determined in 2D
(i.e.\ by angular separation).

Larger values of $S_D$ indicate stronger alignment. So Fig.~\ref{figSD}
indicates that 4-axial LQG PAs are more aligned than expected if they are
randomly drawn from a uniform distribution. Conversely, it also suggests that
2-axial LQG PAs are less aligned (or more orthogonal) than expected. The
physical interpretation of the latter is not straightforward, but likely to
be due to the contribution from orthogonal PAs leading to large dispersion,
and hence a small value of $S_D$.

The approximate empirical standard deviation, and hence the $\pm 1\sigma,
2\sigma$, and $3\sigma$ confidence intervals shown in Fig.~\ref{figSD}, are
valid only for large $n$ while $n_v \ll n$ \citep{Jain2004}. These are not
valid assumptions for much of our range of $S_D$. Therefore, due to this and
the mutual dependence between groups of nearest neighbours, we calculate the
significance level of the S~test using numerical simulations. In
Fig.~\ref{figSL} we show the significance level of the S~test calculated
using 10000 numerical simulations, for values of $S_D$ determined using both
2-axial (green circles) and 4-axial (blue triangles) PAs. Again, this is
shown as a function of nearest neighbours $n_v$.

With 4-axial LQG PAs we test for alignment and orthogonality by combining the
modes, resulting in an alignment only signal (right-hand $S_D$ tail). We find
significance levels generally between 1\% and 5\% for most numbers of nearest
neighbours $n_v$. For most of the $n_v$ range ($30 \le n_v \le 70$) the
significance level is $0.9\% \le \mathrm{SL} \le 2.2\%$, with a mean (median)
of 1.5\% (1.5\%). It is most significant for $n_v \sim 45$, with $\mathrm{SL}
\simeq 0.8\%$.

For 2-axial LQG PAs the orthogonal mode appears to dominate (left-hand $S_D$
tail). We find significance levels generally above 5\% until $n_v \ge
54$. For the remainder of the range of $n_v$ ($54 \le n_v \le 70$) the SL is
$1.9\% \le \mathrm{SL} \le 6.2\%$, with a mean (median) of 3.3\% (3.4\%).

\subsubsection{Typical angular and proper separations}

The number of LQG nearest neighbours $n_v$ is a free parameter explored by
the S~test. We identify these neighbours using the three-dimensional proper
positions of each LQG centroid. The parameter $n_v$ is related to the scale
of the nearest neighbour groups, but because LQGs are not homogeneously
distributed we cannot directly interpret it as corresponding to a particular
scale. We evaluate the relationship between the parameter $n_v$ and the
typical scale of the nearest neighbour groups, defined as the median
separation between each LQG and its $n_v$ nearest neighbours.

Using these relationships, we evaluate the S~test significance levels as a
function of typical separation instead of nearest neighbours $n_v$. The
functions do not differ significantly in shape, because the relationships are
generally linear when $n_v \gtrsim 20$. We find that the correlation is most
significant for typical angular (proper) separations of $\sim 30^{\circ}$
(1.6 Gpc).

\section{Discussion and conclusions}
\label{secConclusions}

We find that LQG PAs are unlikely to be drawn from a uniform distribution
($p$-values $0.008 \lesssim p \lesssim 0.07$). However, similar
non-uniformity is not found in mock LQG catalogues, indicating the LQG
correlation is not found in cosmological simulations. Further, the LQG PA
distribution is bimodal, with modes for weighted PAs at
$\bar{\theta}\sim52\pm2^{\circ}, 137\pm3^{\circ}$ (97\% confidence). This
bimodality is robust to parallel transport destination, with the median
location of the peaks at all 71 LQG locations of
$\bar{\theta}\sim45\pm2^{\circ}, 136\pm2^{\circ}$. These angles are
remarkably close to the mean angles of radio quasar polarisation of
$\bar{\theta}\simeq42^{\circ}$ and $\bar{\theta}\simeq131^{\circ}$, reported
by \cite{Pelgrims2015}\footnote{Errors on these means were not reported} in
two regions coincident with our LQG sample.

LQGs are aligned and orthogonal across very large scales, with a maximum
significance of $\simeq 0.8\%$ ($2.4\sigma$) for groups of $\sim45$ nearest
neighbours, corresponding to typical angular (proper) separations of $\sim
30^{\circ}$ (1.6 Gpc). The statistical significance of this correlation is
marginal, therefore we cannot exclude it being a chance statistical
anomaly. However, its coincidence with regions of quasar-polarisation
alignment \citep[e.g.][]{Hutsemekers1998,Pelgrims2015}, the link between
quasar polarisation and LQG axes \citep{Hutsemekers2014,Pelgrims2016}, and
the similarity between LQG position angles and the preferred angles of quasar
radio polarisation alignment \citep{Pelgrims2015}, suggest an interesting
result.

We find no indication that boundary effects or selection effects have
influenced these results. Three of the LQGs might be truncated by the RA,
Dec.\ boundaries of DR7QSO: removing them from consideration made no
significant difference. Randomly-generated LQGs with related parameters (same
encompassing circle, same number of members) to the real LQGs did not
reproduce the results.

A plausible mechanism for the correlation of LQG orientations on such large
scales is not obvious. We considered the geometry of the
Universe. \citet{vandeWeygaert2007} uses Voronoi tessellation
\citep{Voronoi1908} to describe the observed cosmic web on $\gtrsim 100$ Mpc
scales \citep[see also][]{Icke1987,vandeWeygaert1994}. We speculated whether
a cellular structure to the Universe, such as Voronoi tessellation, or a more
regular crystalline structure, could cause such an effect.

We also considered primordial anisotropies. \citet{Poltis2010} proposed
cosmic strings as an explanation for quasar-polarisation alignments. They
suggest the decay of these would seed correlated primordial magnetic
fields. However, using the CMB, the possible amplitude of these has been
constrained to less than a few nanoGauss
\citep{Planck2016h}. \citet{Hutsemekers2005} suggest the apparent rotation of
mean optical quasar polarisation angle with redshift may be caused by a
global rotation of the Universe, such as that invoked by \citet{Jaffe2005} to
explain large-scale anisotropies in the CMB data. However, from CMB
temperature and polarisation analysis \citet{Saadeh2016} conclude that
the Universe is neither rotating nor anisotropically stretched.

For the geometric and primordial explanations we considered, it is unclear
how they could translate into our observed position angle
distribution. Further, the primordial explanations have been disfavoured by
observations. The origin of the LQG orientation correlation remains
unexplained.

We found no evidence of $\Lambda$CDM cosmological simulations predicting
correlations between objects on Gpc scales, but this had not been
specifically examined for LQGs. Using mock LQG catalogues
\citep{Marinello2016} we found no evidence of LQG correlation in the Horizon
Run 2 simulation \citep{Kim2011}. This suggests that the cosmic web of the
observed Universe differs on the largest scales to this dark-matter-only
$N$-body simulation. It hints that there could, given the caveats associated
with the simulations, be aspects of the large-scale structure that are not
captured by the power spectrum. Running the LQG finder on other cosmological
simulations would be informative. If the correlation in LQG orientation is
confirmed then perhaps it is an unexpected feature of known physics. If it is
not seen then perhaps something is missing from the simulations
(e.g.\ primordial anisotropies) or it is, after all, a statistical fluke
which coincidentally gives rise to the aligned quasar polarisations.

The LQG orientation correlation we found offers a plausible explanation for
the quasar-polarisation alignments reported by many studies
\citep[e.g.][]{Hutsemekers1998,Hutsemekers2001,Jain2004,Cabanac2005,Tiwari2013,Pelgrims2014,Pelgrims2015,Pelgrims2019}.
If LQG axes are preferentially aligned at $\bar{\theta}\sim45\pm2^{\circ},
136\pm2^{\circ}$ (this work, median modes), and if quasar polarisation
vectors are preferentially parallel and orthogonal to LQG axes
\citep{Hutsemekers2014,Pelgrims2016}, this could result in polarisation
vectors with preferred angles of $\sim42^{\circ}$ and $\sim131^{\circ}$
\citep{Pelgrims2015}. Our results therefore offer corroborating evidence for,
and enhancement of, the intrinsic alignment interpretation of these studies.

Quasar-polarisation alignment is also detected in the south Galactic cap
\citep[SGC, e.g.][]{Hutsemekers1998,Pelgrims2015}, which is not coincident
with our LQG sample in the north Galactic cap (NGC). The forthcoming 4-metre
Multi-Object Spectroscopic Telescope (4MOST) Active Galactic Nuclei survey
\citep{Merloni2019} will survey a million $z \lesssim 2.5$ quasars over $\sim
10000 \ \mathrm{deg}^2$, with first light expected in 2022. This could
deliver an LQG sample in the SGC for similar evaluation to our work in the
NGC. Of particular interest would be whether LQG orientation again
corresponds to the preferred angle of quasar radio polarisation alignment,
which differ between NGC and SGC \citep[e.g.][]{Pelgrims2016thesis}.

Our results are based on a sample of 71 LQGs at redshifts $1.0 \leq z \leq
1.8$, which were detected using the SDSS DR7QSO catalogue
\citep{Schneider2010} of $\sim 105$k quasars across $\sim 7600
\ \mathrm{deg}^2$. Forthcoming spectroscopic surveys will deliver a far
larger sample of quasars, e.g.\ the Dark Energy Spectroscopic Instrument
\citep[DESI,][]{DESI2016} 5-year survey aims to target 1.7 million $z < 2.1$
quasars covering $\sim 14000 \ \mathrm{deg}^2$, beginning in May 2021. This
has the potential to deliver a larger sample of LQGs for a better assessment
of their correlation.

If the LQG orientation correlation is real, it represents large-scale
structure alignment over $\gtrsim$ Gpc scales, larger than those predicted by
cosmological simulations and at least an order of magnitude larger than any
so far observed, with the exception of quasar-polarisation / radio-jet
alignment. Careful statistical analysis is required before making inferences
about whether such a large-scale correlation challenges the assumption of
large-scale statistical isotropy and homogeneity of the Universe.

To conclude, we find large-scale correlation of LQG orientations, which we
report here for the first time. This helps explain a substantial body of work
on quasar-polarisation / radio-jet alignment, but at the expense of raising
potentially even more challenging questions about the origin of the LQG
correlation and its implications for isotropy and homogeneity. Forthcoming
surveys and the other future work we suggest here will illuminate LQGs and
their intriguing correlation further.

\section*{Acknowledgements}

We thank Srinivasan Raghunathan for many helpful discussions. TF acknowledges
receipt of a STFC PhD studentship. We thank the referee for thoughtful and
helpful comments.

\section*{Data availability statement}

The datasets were derived from sources in the public domain: {\small
  https://classic.sdss.org/dr7/products/value\_added/qsocat\_dr7.html}.




\bibliographystyle{mnras}
\bibliography{bibliography} 




\clearpage

\appendix

\section{LQG finder}
\label{appLQGfinder}

\citet{Clowes2012,Clowes2013} detect LQG candidates using a three-dimensional
single-linkage hierarchical clustering algorithm, also known as
friends-of-friends (FoF). This is equivalent to a three-dimensional minimal
spanning tree (MST). These type of methods are widely used to detect galaxy
clusters, superclusters, voids, and filaments
\citep[e.g.][]{Press1982,Einasto1997,Park2012,Pereyra2020}, as well as LQGs
\citep[e.g.][]{Clowes2012,Clowes2013,Nadathur2013,Einasto2014,Park2015}. They
make no assumptions about cluster morphology.

Using the terminology of \citet{Barrow1985}, MST treats the dataset as a
graph made up of vertices (nodes, in this case quasars) which are connected
by edges (straight lines). This is a `tree' when it has no closed paths
(sequence of edges) and a `spanning' tree when it contains all the
vertices. For any graph, there are multiple possible spanning trees; the
`minimal' spanning tree is that of minimal length (sum of edge lengths). To
identify clusters within the MST it may be separated, where edges exceeding a
certain length are removed, leaving groups of objects with mutual separations
less than this `linkage length'.

The choice of linkage length is crucial - too long and clusters merge to fill
the entire volume, too short they break up into pairs and triplets
\citep{Graham1995}. \citet{Pilipenko2007} categorises the criteria for making
this choice as physical or formal. With a priori knowledge of physical
parameters (e.g.\ size, membership, density) of the clusters, it is possible
to choose the scale that maximises the fraction of clusters with those
parameters. The criterion used by \citet{Graham1995} is an example of a
physical approach; they choose the scale that maximises the number of
clusters of a minimum membership. An example formal approach would be to
choose a scale based on the mean nearest-neighbour separation.

\citet{Clowes2012,Clowes2013} use this latter approach. Their quasar sample
has a mean nearest-neighbour separation of $\sim 74 \ \mathrm{Mpc}$. They
also account for uncertainties in their edge lengths (i.e.\ proper
distances) due to redshift errors and peculiar velocities \citep[for
  estimates, see][]{Clowes2012} and choose a linkage length of $100
\ \mathrm{Mpc}$.

\citet{Clowes2012,Clowes2013} estimate the overdensity and statistical
significance of their LQGs using a convex hull of member spheres (CHMS)
method. This is described in detail in \citet{Clowes2012}, but briefly, for
each LQG they calculate the volume of a convex hull of spheres of radius half
the mean edge length at each vertex (member quasar location). The CHMS volume
of an LQG of $m$ members is then compared with the distribution of CHMS
volumes of clusters of $m$ points in Monte Carlo simulations of the same size
and density as their control area, in order to estimate overdensity and
statistical significance. We note this method of estimating statistical
significance is not universally accepted
\citep{Nadathur2013,Pilipenko2013,Park2015}.

The Huge-LQG and Clowes-Campusano LQG, amongst others, have been
independently detected using different FoF algorithms
\citep{Nadathur2013,Einasto2014,Park2015}. Indeed, and unsurprisingly, our
LQG sample has many objects in common with the publicly available
catalogue\footnote{\url{https://vizier.u-strasbg.fr/viz-bin/VizieR?-source=J/A+A/568/A46}}
of \citet{Einasto2014}.  \citet{Einasto2014} appear to have followed closely
the approach of \citet{Clowes2012,Clowes2013} in terms of input data,
selection algorithm, linkage scale, and cosmological model and parameters,
but applied to a reduced area of DR7QSO. \citet{Einasto2014} differed
substantially, however, in not providing any measures of statistical
significance or overdensity, which are important for ranking the LQG
candidates.

\section{LQG sample}
\label{appLQGsample}

Fig.~\ref{figAllLQGsPt1} illustrates the 71 LQGs in our sample, shown in
tangent plane projection.

\begin{figure} 
    \centering
    \includegraphics[width=0.8\columnwidth]{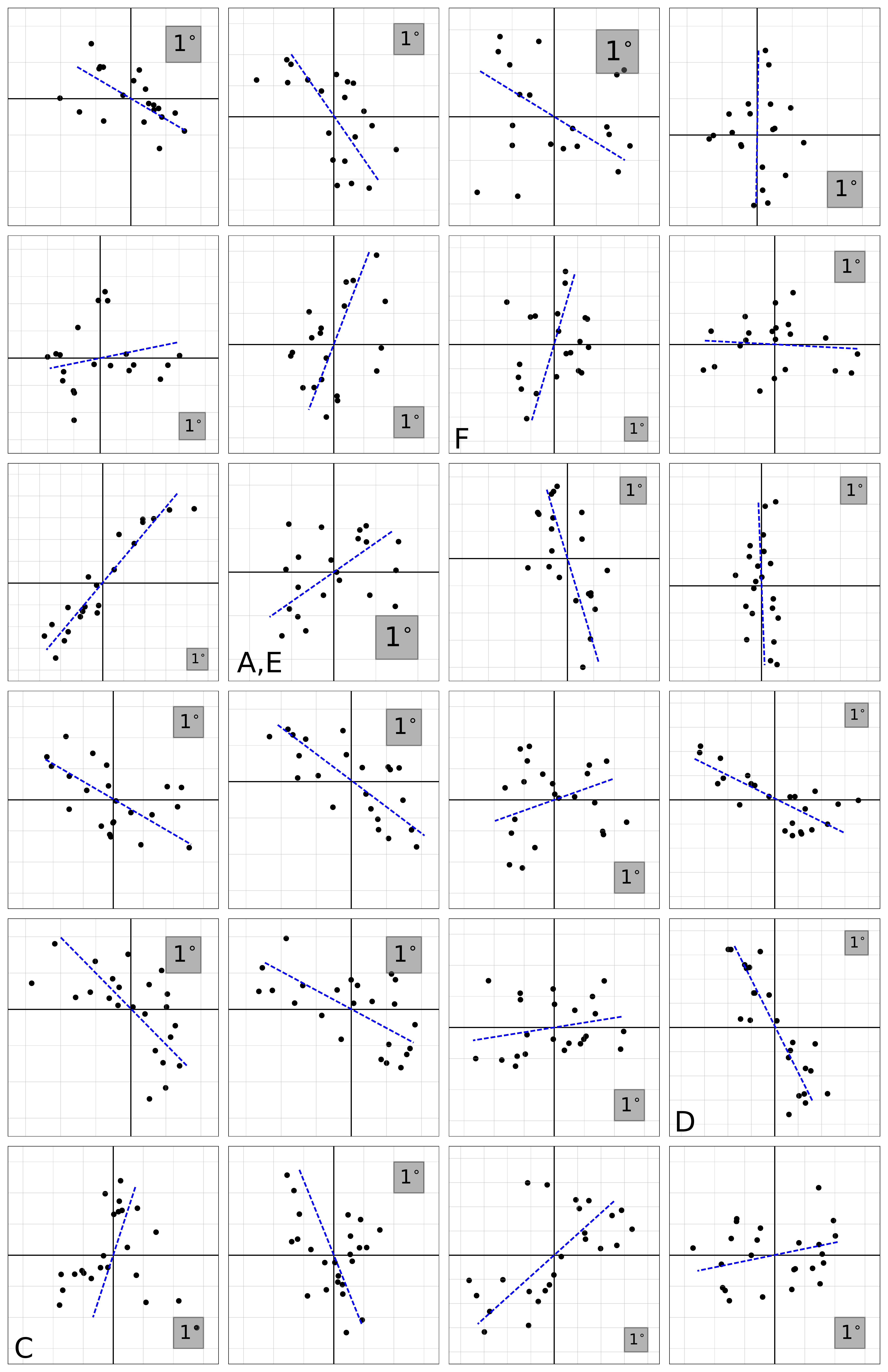}
    \caption[Tangent plane projection of the 71 LQGs in our sample]{(a)
      Tangent plane projection of 1-24 (across then down) of the 71 LQGs in
      our sample, in Cartesian coordinates. Member quasars shown as black
      dots, orthogonal distance regression fit shown as dashed blue
      line. Solid black lines indicate $x=0$ and $y=0$, and grey square
      illustrates scale ($1^{\circ}$). LQGs labelled A, C (D, E, F) are
      discussed in Fig.~\ref{figPA2D3D} (Fig.~\ref{figBootstraps}).}
    \label{figAllLQGsPt1}
\end{figure}

\begin{figure}
    \ContinuedFloat
    \centering
    \includegraphics[width=0.8\columnwidth]{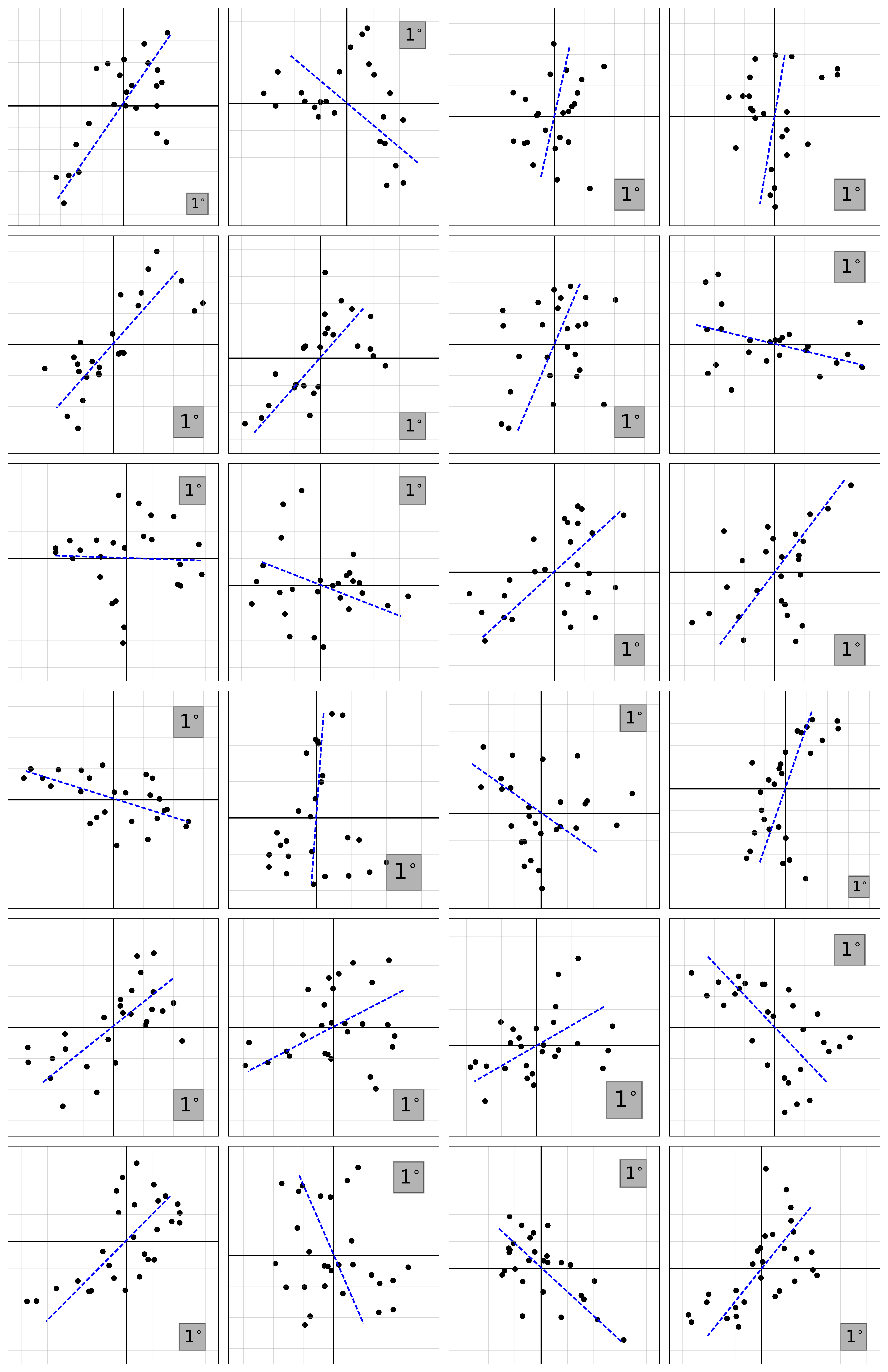}
    \caption[]{(b) Tangent plane projection of 25-48 (across then down) of
      the 71 LQGs in our sample, in Cartesian coordinates. Other plot
      details are as for Fig.~\ref{figAllLQGsPt1} panel (a).}
    \label{figAllLQGsPt2}
\end{figure}

\begin{figure}
    \ContinuedFloat
    \centering
    \includegraphics[width=0.8\columnwidth]{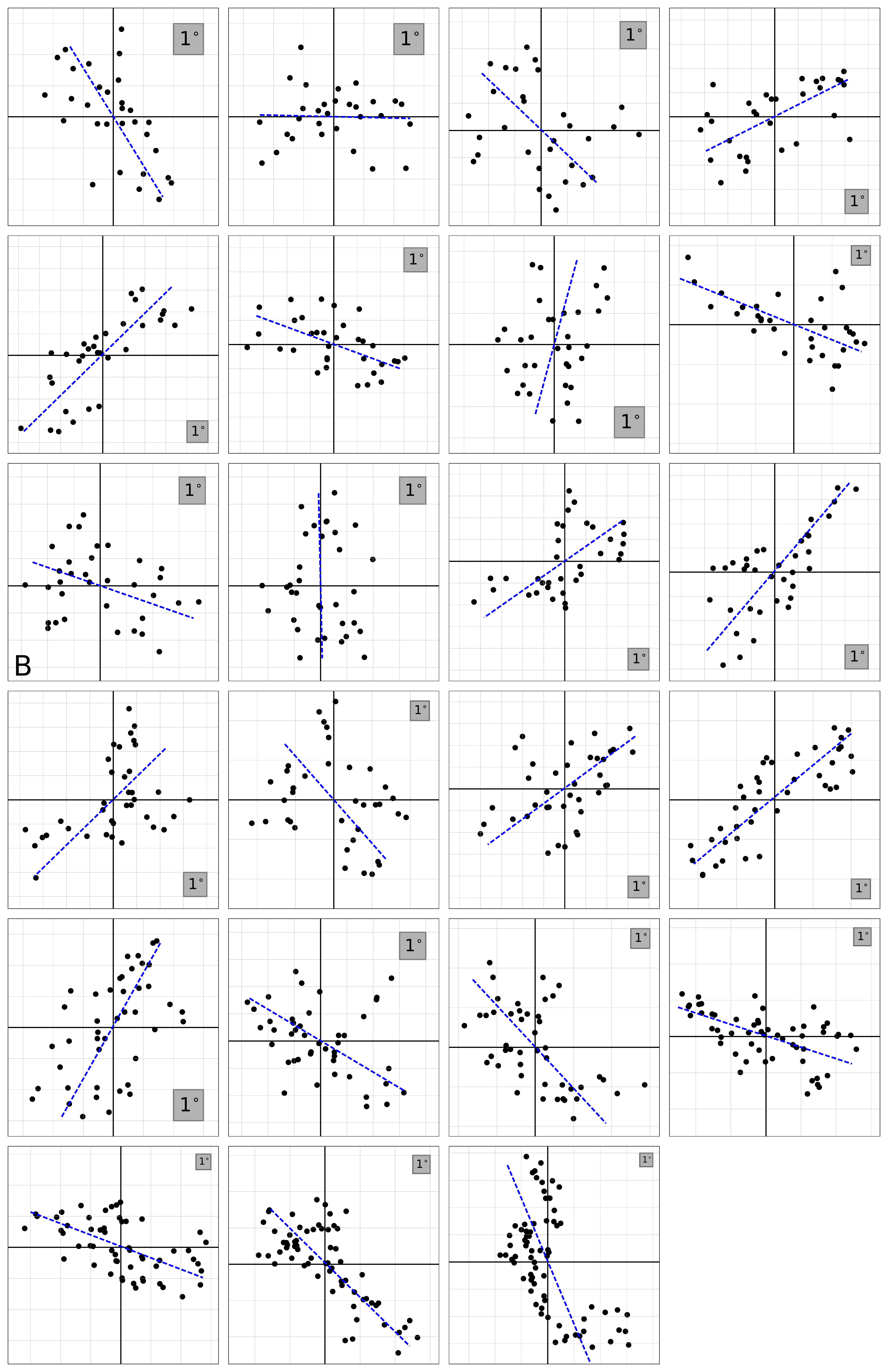}
    \caption[]{(c) Tangent plane projection of 49-71 (across then down) of
      the 71 LQGs in our sample, in Cartesian coordinates. Other plot
      details are as for Fig.~\ref{figAllLQGsPt1} panel (a).
      LQG labelled B is discussed in Fig.~\ref{figPA2D3D}.}
    \label{figAllLQGsPt3}
\end{figure}

\section{LQG orientation}
\label{appLQGorientation}

\subsection{LQG position angles --- presentation}
\label{subappLQGorientation_pres}

The position angle (PA) of a large quasar group can be calculated in either
two or three dimensions. In two dimensions we treat the quasars as points on
the celestial sphere, whereas in three dimensions we take into account their
proper radial distances.

\begin{figure} 
    \centering
    \includegraphics[width=0.7\columnwidth]{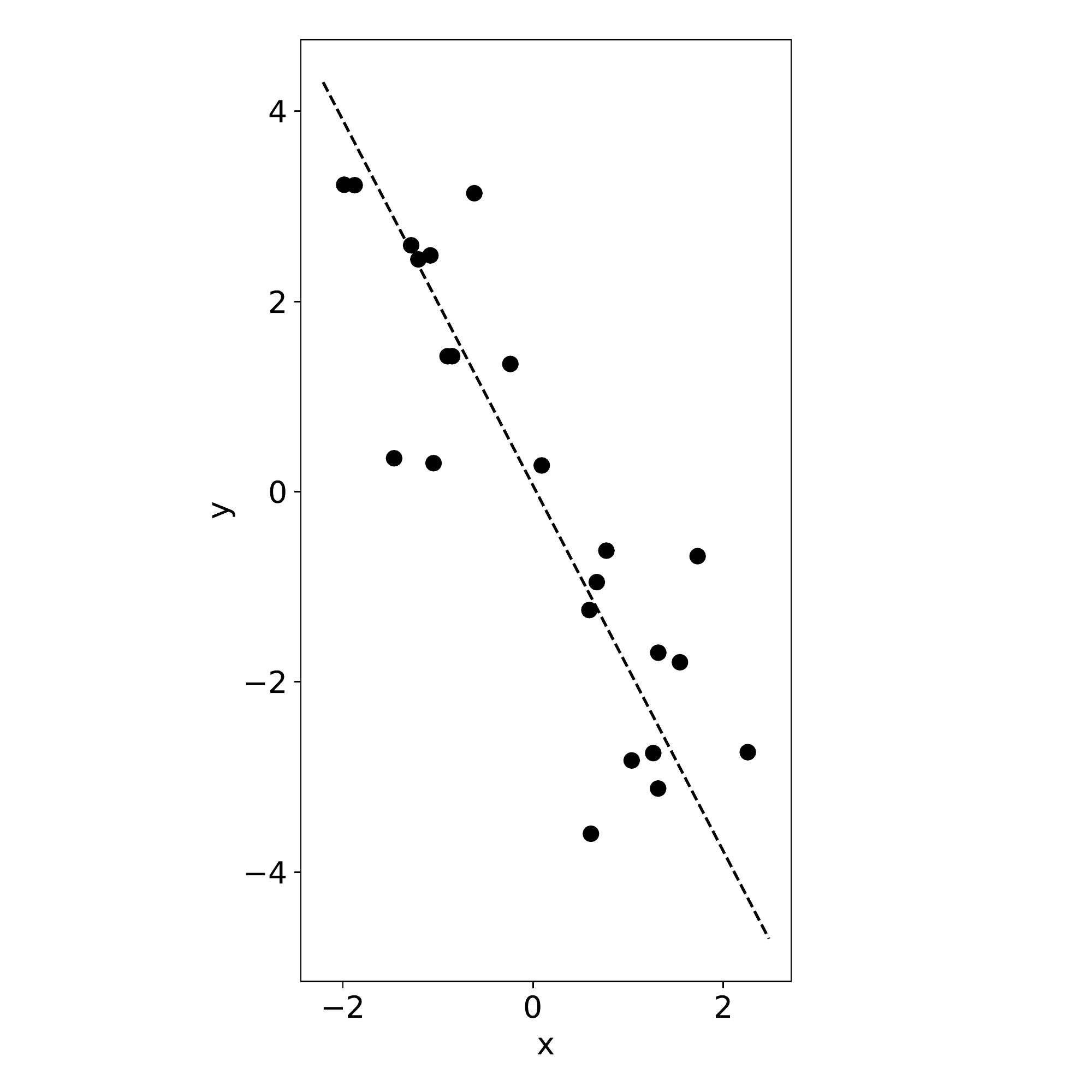}
    \caption[Example position angle calculated by 2D approach]{Tangent plane
      projection of an LQG, with coordinates ($\alpha$, $\delta$) projected
      to ($x$, $y$). Member quasars are shown as black dots and orthogonal
      distance regression fit shown as dashed line. Axes are labelled in
      degrees. This LQG is also shown in Figs.~\ref{figAllLQGsPt1} and
      \ref{figBootstraps}, labelled D. Using the 2D PA approach, PA =
      $152.5^{\circ}$.}
    \label{fig2DPA}
\end{figure}

For the 2D approach quasar positions (in right ascension and declination) are
projected onto the tangent plane as Cartesian $(x,y)$ points. This plane
meets the celestial sphere at the centre of gravity of the LQG $(x,y) =
(0,0)$, calculated assuming quasars are point-like unit masses. To determine
LQG orientation we use orthogonal distance regression (ODR) of these
projected points (see Fig.~\ref{fig2DPA} for an example). This minimises the
sum of the squares of the orthogonal residuals between the points and the
line \citep[for a discussion of OLS, ODR, and other regression methods,
  see][]{Isobe1990}.

\begin{figure}
    \centering
    \includegraphics[width=0.8\columnwidth]{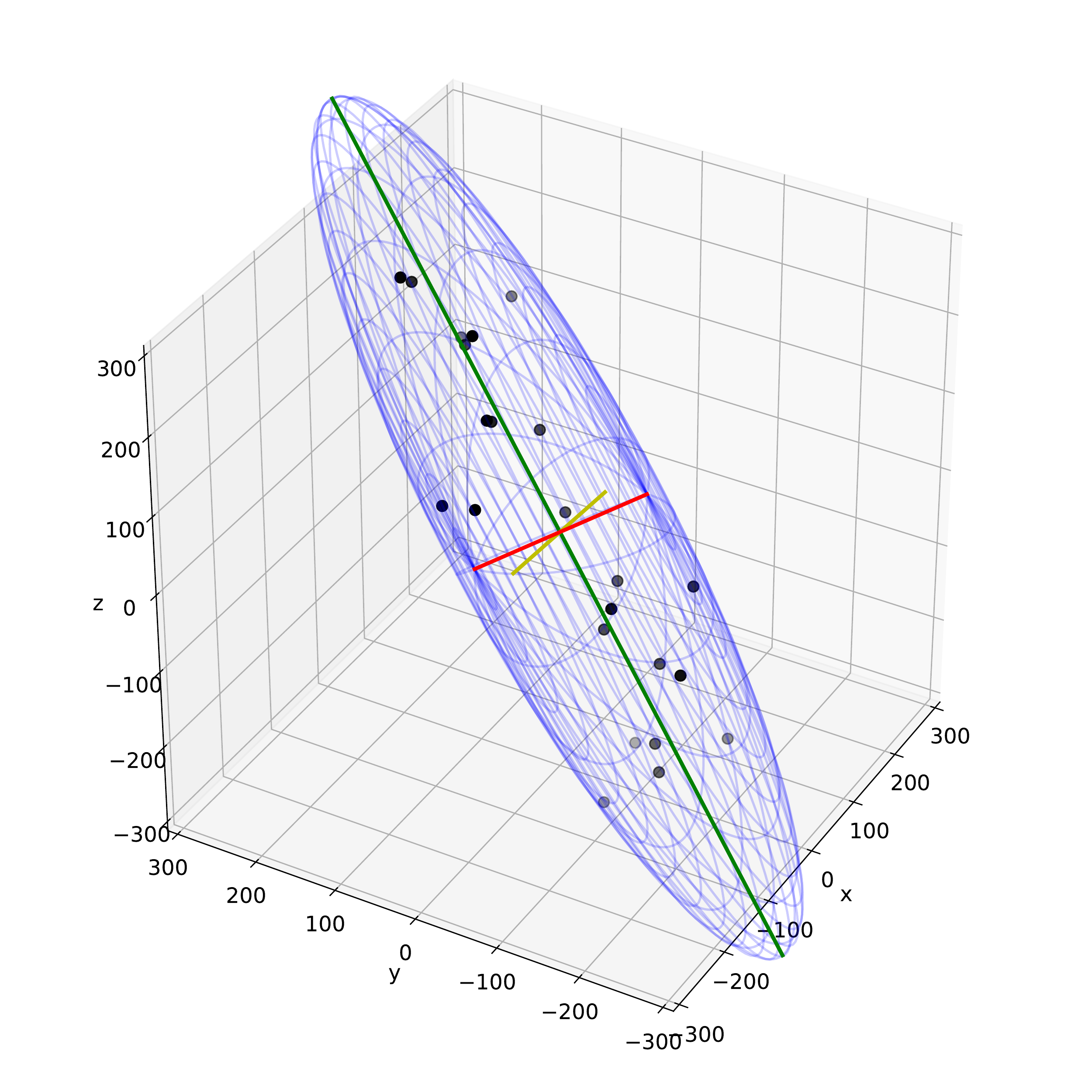}
    \caption[Example position angle calculated by 3D approach]{Same LQG as
      Fig.~\ref{fig2DPA}, but shown in 3D proper coordinates oriented with
      line-of-sight orthogonal to the page. Axes of enclosing ellipsoid
      (green = \textit{a}-axis, yellow = \textit{b}-axis, red =
      \textit{c}-axis) constructed from eigenvectors and eigenvalues. Axes
      are labelled in Mpc. Using the 3D PA approach, 2D projected PA =
      $152.3^{\circ}$.}
    \label{fig3DPA}
\end{figure}

For the 3D approach the covariance matrix of the quasar proper positions is
decomposed into its eigenvectors and eigenvalues. Again, quasars are assumed
to be point-like unit masses. The axes of a confidence ellipsoid
(e.g.\ Fig.~\ref{fig3DPA}) are constructed from the eigenvectors and
eigenvalues; each axis is in the direction of its eigenvector and its length
$\ell$ is a function of its eigenvalue $\lambda$, specifically $\ell \propto
\sqrt{\lambda}$. The first principal component (ellipsoid major axis) is
given by the eigenvector with the largest eigenvalue. Finally, this is
projected onto the plane orthogonal to the line-of-sight to define the (2D
projected) PA.

PAs determined by the two approaches may differ, for example the 2D approach
may be susceptible to projection effects and the 3D approach may be
susceptible to redshift-space distortions. The orientation of the LQG with
respect to the line-of-sight and its morphology may also induce
differences. We expect PAs determined by the two approaches to agree well
when the LQG is linear and orthogonal to the line-of-sight, but they may
differ significantly when the LQG is broad, crooked, curved or aligned along
the line-of-sight.

\begin{figure} 
    \centering
	\includegraphics[width=0.7\columnwidth]{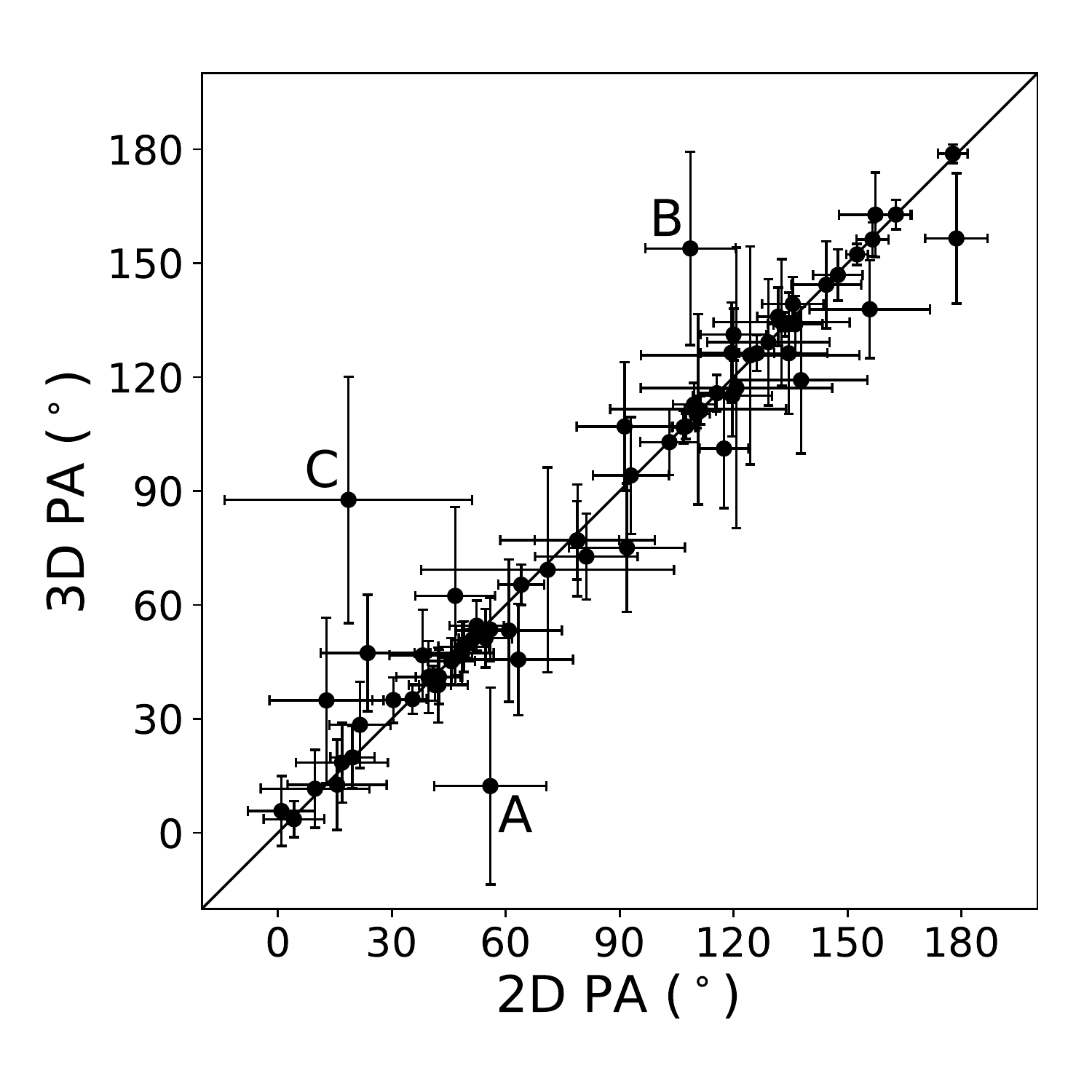}
	\caption[Position angles calculated by 2D and 3D approaches
          (comparison)]{Position angles of 71 LQGs calculated by 2D and 3D
          approaches. The two generally agree well, with the three widest
          outliers (A, B, C) due to the geometry of those particular LQGs
          (see text and Fig.~\ref{figAllLQGsPt1}). Error bars show half-width
          confidence intervals estimated using bootstrap re-sampling.}
	\label{figPA2D3D}
\end{figure}

For our sample of 71 LQGs we find that the two approaches are generally
consistent. Fig.~\ref{figPA2D3D} shows the PAs calculated using both the 2D
and 3D approaches. Note that the PAs have not been parallel transported, but
these angles serve as a useful comparison between the two approaches. The
error bars are the half-width confidence intervals estimated using bootstrap
re-sampling. Note that, usually, the measurement uncertainties are slightly
larger for the 3D approach.

The three widest outliers from the 1:1 diagonal line in Fig.~\ref{figPA2D3D}
are due to the geometry of these particular LQGs (A, B, C; also labelled in
Fig.~\ref{figAllLQGsPt1}). Two of these LQGs (A and B) have their major axes
oriented towards the line-of-sight, and not significantly longer than their
first minor axes. Indeed, for both of these, the 2D approach fits a
regression comparable to the first minor axes rather than the major axes. One
LQG (C) is very irregular so linear fits are poor, and corresponding PAs are
uncertain, in both the 2D and 3D approaches. The PAs of all three of these
LQGs are given little weight by goodness-of-fit weighting
(section~\ref{subLQGOrientations}).

Both the 2D and 3D approaches have been used to determine the PAs of
LQGs. \citet{Hutsemekers2014} use the 2D approach to demonstrate alignment of
quasars' optical linear polarisation with LQG axes, while
\citet{Pelgrims2016} use the 3D approach to evidence alignment of quasars'
radio polarisation with more LQG axes. The latter derived eigenvectors and
eigenvalues from the inertia tensor rather than covariance matrix; results
are equivalent. \citet{Pelgrims2016} report that for the 2D approach PAs
calculated using ODR are consistent (within $1^{\circ}$) with those
determined using the inertia tensor.

\citet{Pelgrims2016} show that both approaches usually agree well, and argue
that the 3D approach is more physically motivated. We agree, but note that
any 3D analysis is susceptible to redshift errors. The quasars in our sample
are from SDSS DR7QSO, which typically has quoted redshift errors of $\Delta z
\sim 0.004$ \citep{Schneider2010}. There is also evidence for systematic
errors of $\Delta z \sim 0.003$ \citep{Hewett2010}. Using Monte Carlo
simulations we find that these redshift errors introduce uncertainty in the
PA, generally of a few degrees, but up to $\sim 30^{\circ}$ for LQGs
particularly oriented along the line-of-sight (e.g.\ LQGs A and B).

Furthermore, in the 3D approach, there is also the potential for errors due
to redshift-space distortions from the quasars' peculiar velocities, causing
their real-space distribution to be either elongated \citep{Jackson1972} or
squashed \citep{Kaiser1987} along the line-of-sight. We find that the
measurement uncertainties (Appendix~\ref{appPAuncertainties}) are slightly
larger for the 3D approach. We therefore base our analysis on the
two-dimensional approach; tangent plane projection of the LQG and orthogonal
distance regression of the projected quasars.

\subsection{LQG position angles --- tabulation}
\label{subappLQGorientation_tab}

Table~\ref{tabLQGDataFull} presents the results of both the two-dimensional
and three-dimensional approaches to determining position angle. To recap, the
2D approach involves tangent plane projection of the LQG quasars, followed by
orthogonal distance regression (ODR) of the projected points. The 3D approach
requires determining the proper coordinates of the LQG quasars, performing
principal component analysis on the covariance matrix of these, then tangent
plane projection of the resultant major axis.

The results of both approaches generally agree well. For both approaches,
bootstrap re-sampling with replacement is used to estimate the uncertainty in
the form of the half-width confidence interval (HWCI, $\gamma_h$) of 10000
bootstraps. The PAs listed in Table~\ref{tabLQGDataFull} are measured in situ
at the location of each LQG, and will have parallel transport corrections
applied before statistical analysis.


\begin{table*}
    \centering
    \caption[Large quasar group locations and position angles (full LQG
      sample)]{The 71 large quasar groups, where $m$ is the number of
      members, and $\bar{\alpha}$, $\bar{\delta}$, and $\bar{z}$ are the mean
      right ascension, declination, and redshift of the member quasars. The
      normalized goodness-of-fit weight $w$ (Eq.~\ref{eqWeight}) is scaled by
      $w_{71} = w \times 71$ for clarity, and to distinguish those LQGs
      weighted higher ($w_{71} > 1$) or lower ($w_{71} < 1$) than the mean
      $\bar{w}$. Position angle $\theta$ and half-width confidence interval
      $\gamma_h$ are shown for both the 2D and 3D approaches. The ratio of
      LQG ellipsoid axes lengths (from the 3D approach,
      Appendix~\ref{appLQGorientation}) is given by $a:b:c$. This list was
      summarized in Table~\ref{tabLQGDataPart}.}
    \label{tabLQGDataFull}
    \begin{tabular}{crrrrrrrrc}
        \hline
        & \multicolumn{2}{r}{J2000 ($^\circ$)} &&& \multicolumn{2}{r}{2D PA ($^\circ$)} & \multicolumn{2}{r}{3D PA ($^\circ$)} & \\
        $m$ & $\bar{\alpha}$ & $\bar{\delta}$ & $\bar{z}$ & $w_{71}$ & $\theta$ & $\gamma_h$ & $\theta$ & $\gamma_h$ & $a:b:c$ \\
        \hline
        20 & 121.1 & 27.9 & 1.73 & 1.13 & 119.7 & 10.4 & 115.1 & 10.7 & 0.50:0.30:0.21 \\
        20 & 151.5 & 48.6 & 1.46 & 1.21 & 144.4 & 9.2 & 144.3 & 11.5 & 0.50:0.33:0.17 \\
        20 & 155.9 & 12.8 & 1.50 & 0.32 & 120.7 & 25.2 & 117.2 & 37.0 & 0.44:0.43:0.14 \\
        20 & 163.6 & 16.9 & 1.57 & 1.00 & 0.9 & 8.8 & 5.7 & 9.2 & 0.47:0.33:0.20 \\
        20 & 178.0 & 1.2 & 1.23 & 0.37 & 78.9 & 20.3 & 77.0 & 14.7 & 0.51:0.31:0.17 \\
        20 & 216.4 & 1.4 & 1.11 & 0.86 & 21.6 & 8.1 & 28.4 & 11.3 & 0.49:0.30:0.21 \\
        21 & 133.5 & 41.2 & 1.40 & 0.72 & 16.9 & 12.1 & 18.5 & 10.6 & 0.49:0.33:0.18 \\
        21 & 170.6 & 16.8 & 1.07 & 0.90 & 93.0 & 10.0 & 94.1 & 15.4 & 0.47:0.37:0.17 \\
        21 & 191.8 & 11.0 & 1.06 & 4.43 & 40.9 & 2.8 & 41.0 & 2.9 & 0.66:0.24:0.10 \\
        21 & 209.1 & 3.2 & 1.56 & 0.77 & 55.9 & 14.8 & 12.3 & 25.9 & 0.49:0.30:0.21 \\
        21 & 212.9 & 12.6 & 1.55 & 1.72 & 162.7 & 3.9 & 162.8 & 3.9 & 0.63:0.22:0.15 \\
        21 & 231.2 & 25.2 & 1.51 & 4.57 & 177.7 & 3.9 & 178.8 & 2.4 & 0.57:0.29:0.14 \\
        22 & 136.8 & 49.5 & 1.19 & 1.03 & 119.4 & 8.2 & 126.4 & 13.2 & 0.46:0.38:0.17 \\
        22 & 182.0 & 55.5 & 1.70 & 1.88 & 126.1 & 4.6 & 126.3 & 4.7 & 0.56:0.23:0.20 \\
        22 & 217.8 & -0.8 & 1.31 & 0.42 & 71.0 & 33.3 & 69.2 & 27.0 & 0.42:0.33:0.25 \\
        23 & 155.9 & 53.5 & 1.48 & 1.79 & 115.6 & 4.8 & 115.8 & 4.8 & 0.60:0.22:0.19 \\
        23 & 166.4 & 37.1 & 1.31 & 0.97 & 134.5 & 10.2 & 126.3 & 15.9 & 0.44:0.37:0.19 \\
        23 & 171.3 & 14.0 & 1.20 & 1.57 & 117.5 & 6.4 & 101.2 & 15.7 & 0.48:0.37:0.15 \\
        23 & 180.5 & 6.0 & 1.29 & 1.26 & 81.2 & 13.5 & 72.8 & 11.3 & 0.57:0.25:0.18 \\
        23 & 209.5 & 34.3 & 1.65 & 1.99 & 152.5 & 2.8 & 152.3 & 2.8 & 0.68:0.17:0.15 \\
        23 & 214.3 & 31.8 & 1.48 & 0.27 & 18.6 & 32.6 & 87.7 & 32.5 & 0.47:0.35:0.18 \\
        24 & 119.1 & 18.6 & 1.28 & 1.06 & 157.3 & 9.6 & 162.8 & 11.1 & 0.46:0.32:0.23 \\
        24 & 139.6 & 2.5 & 1.20 & 0.61 & 48.9 & 6.7 & 49.0 & 6.7 & 0.58:0.29:0.13 \\
        24 & 171.1 & 17.6 & 1.52 & 0.83 & 78.7 & 11.2 & 77.0 & 10.3 & 0.50:0.28:0.22 \\
        24 & 179.6 & 65.0 & 1.08 & 0.71 & 35.4 & 3.7 & 35.2 & 3.9 & 0.55:0.23:0.22 \\
        24 & 205.0 & 12.0 & 1.36 & 0.61 & 129.2 & 16.1 & 129.2 & 16.6 & 0.46:0.33:0.20 \\
        24 & 217.1 & 33.8 & 1.11 & 0.99 & 12.8 & 15.0 & 34.9 & 21.8 & 0.54:0.28:0.18 \\
        24 & 217.1 & 57.5 & 1.70 & 0.75 & 9.8 & 14.3 & 11.6 & 10.3 & 0.52:0.28:0.20 \\
        25 & 142.5 & 31.3 & 1.33 & 1.70 & 42.4 & 6.1 & 41.1 & 7.2 & 0.50:0.34:0.15 \\
        25 & 149.5 & 43.7 & 1.16 & 0.81 & 42.3 & 7.8 & 38.9 & 9.9 & 0.45:0.35:0.20 \\
        25 & 186.0 & 3.6 & 1.19 & 0.61 & 23.6 & 12.3 & 47.3 & 15.3 & 0.48:0.32:0.20 \\
        25 & 196.7 & 36.5 & 1.48 & 0.64 & 103.1 & 7.7 & 102.9 & 8.6 & 0.47:0.33:0.20 \\
        25 & 231.9 & 43.6 & 1.20 & 0.30 & 91.9 & 15.3 & 75.1 & 16.9 & 0.44:0.36:0.21 \\
        26 & 160.3 & 53.5 & 1.18 & 0.33 & 110.6 & 23.2 & 111.5 & 25.1 & 0.38:0.34:0.28 \\
        26 & 171.7 & 24.2 & 1.10 & 0.78 & 48.5 & 8.3 & 47.3 & 8.1 & 0.46:0.29:0.24 \\
        26 & 206.0 & 6.5 & 1.43 & 0.75 & 38.1 & 8.7 & 46.7 & 12.0 & 0.46:0.36:0.18 \\
        26 & 224.2 & 61.6 & 1.57 & 1.52 & 106.8 & 3.1 & 106.8 & 4.3 & 0.51:0.33:0.16 \\
        26 & 245.6 & 40.3 & 1.55 & 0.58 & 4.2 & 8.0 & 3.5 & 4.7 & 0.49:0.26:0.25 \\
        27 & 184.1 & 52.7 & 1.18 & 0.37 & 124.4 & 28.7 & 125.7 & 28.7 & 0.40:0.35:0.25 \\
        27 & 204.0 & 13.7 & 1.16 & 0.98 & 19.6 & 5.8 & 19.9 & 8.2 & 0.50:0.36:0.13 \\
        27 & 221.8 & 52.4 & 1.55 & 1.12 & 52.3 & 7.2 & 54.5 & 6.6 & 0.54:0.26:0.20 \\
        27 & 225.5 & 57.3 & 1.52 & 0.54 & 63.3 & 14.4 & 45.6 & 14.6 & 0.45:0.37:0.18 \\
        27 & 231.2 & 15.4 & 1.56 & 1.01 & 60.8 & 14.0 & 53.2 & 18.7 & 0.43:0.35:0.22 \\
        28 & 138.6 & 14.8 & 1.54 & 1.22 & 135.6 & 8.1 & 139.3 & 7.1 & 0.63:0.21:0.15 \\
        28 & 177.0 & 43.1 & 1.54 & 1.10 & 45.7 & 6.2 & 45.2 & 6.1 & 0.52:0.25:0.23 \\
        28 & 192.3 & 10.7 & 1.36 & 0.51 & 155.8 & 15.8 & 137.9 & 12.9 & 0.49:0.32:0.19 \\
        28 & 212.0 & 61.7 & 1.15 & 1.33 & 131.7 & 5.5 & 135.9 & 7.6 & 0.61:0.23:0.16 \\
        30 & 138.1 & 7.9 & 1.56 & 0.85 & 39.7 & 8.4 & 41.0 & 9.5 & 0.51:0.32:0.17 \\
        30 & 217.2 & 10.5 & 1.67 & 0.99 & 147.5 & 6.5 & 146.9 & 6.8 & 0.56:0.23:0.21 \\
        30 & 234.0 & 21.8 & 1.60 & 0.61 & 91.3 & 12.6 & 107.0 & 16.9 & 0.46:0.33:0.21 \\
        31 & 176.5 & 38.3 & 1.28 & 0.37 & 132.6 & 17.9 & 134.4 & 16.6 & 0.46:0.34:0.19 \\
        31 & 202.7 & 21.3 & 1.08 & 0.74 & 64.1 & 6.0 & 65.3 & 5.3 & 0.56:0.24:0.20 \\
        32 & 177.4 & 32.7 & 1.18 & 2.11 & 46.6 & 4.5 & 46.1 & 4.6 & 0.59:0.23:0.18 \\
        33 & 126.6 & 20.3 & 1.44 & 1.15 & 109.6 & 5.6 & 112.8 & 5.7 & 0.52:0.29:0.20 \\
        33 & 157.8 & 20.7 & 1.59 & 1.07 & 15.6 & 13.0 & 12.7 & 11.9 & 0.41:0.33:0.26 \\
        33 & 183.6 & 11.0 & 1.08 & 0.69 & 111.3 & 4.1 & 111.2 & 3.7 & 0.54:0.25:0.21 \\
        \hline
    \end{tabular}
\end{table*}

\begin{table*}
    \centering
    \caption*{continued...}
    \label{tabLQGDataFullExt}
    \begin{tabular}{crrrrrrrrc}
        \hline
        & \multicolumn{2}{r}{J2000 ($^\circ$)} &&& \multicolumn{2}{r}{2D PA ($^\circ$)} & \multicolumn{2}{r}{3D PA ($^\circ$)} & \\
        $m$ & $\bar{\alpha}$ & $\bar{\delta}$ & $\bar{z}$ & $w_{71}$ & $\theta$ & $\gamma_h$ & $\theta$ & $\gamma_h$ & $a:b:c$ \\
        \hline
        34 & 162.3 & 5.3 & 1.28 & 0.60 & 108.6 & 11.9 & 153.9 & 25.5 & 0.44:0.37:0.19 \\
        34 & 228.5 & 8.1 & 1.22 & 0.71 & 178.7 & 8.2 & 156.5 & 17.2 & 0.45:0.34:0.21 \\
        34 & 234.5 & 10.7 & 1.24 & 0.88 & 55.9 & 7.4 & 53.5 & 8.3 & 0.48:0.29:0.24 \\
        36 & 189.0 & 44.1 & 1.39 & 1.22 & 41.3 & 4.2 & 38.9 & 4.1 & 0.53:0.28:0.19 \\
        37 & 189.5 & 20.3 & 1.46 & 0.66 & 46.7 & 10.5 & 62.4 & 23.4 & 0.42:0.39:0.19 \\
        38 & 161.6 & 3.5 & 1.11 & 0.35 & 137.7 & 17.5 & 119.2 & 19.3 & 0.46:0.37:0.18 \\
        38 & 227.6 & 41.4 & 1.54 & 0.50 & 54.7 & 7.0 & 51.2 & 7.7 & 0.50:0.34:0.16 \\
        41 & 205.3 & 50.4 & 1.39 & 1.36 & 51.3 & 2.7 & 51.0 & 2.7 & 0.63:0.22:0.15 \\
        43 & 231.0 & 47.8 & 1.57 & 0.76 & 30.4 & 5.5 & 35.0 & 6.0 & 0.47:0.32:0.20 \\
        44 & 208.7 & 25.8 & 1.28 & 0.45 & 120.0 & 8.6 & 131.2 & 6.8 & 0.54:0.27:0.19 \\
        46 & 226.7 & 16.7 & 1.09 & 0.63 & 136.2 & 7.2 & 133.9 & 7.5 & 0.47:0.30:0.23 \\
        55 & 196.5 & 27.1 & 1.59 & 0.95 & 107.5 & 3.4 & 107.0 & 3.4 & 0.58:0.24:0.18 \\
        56 & 167.0 & 33.8 & 1.11 & 0.81 & 110.2 & 3.5 & 110.4 & 3.8 & 0.50:0.29:0.21 \\
        64 & 196.4 & 39.9 & 1.14 & 0.83 & 133.6 & 3.0 & 133.9 & 3.2 & 0.48:0.36:0.17 \\
        73 & 164.1 & 14.1 & 1.27 & 0.76 & 156.6 & 4.2 & 156.3 & 4.5 & 0.55:0.28:0.16 \\
        \hline
    \end{tabular}
\end{table*}

\section{Axial data}
\label{appAxial}

Position angle data are axial [$0^{\circ}$, $180^{\circ}$), more specifically
  2-axial; $0^{\circ}$ and $180^{\circ}$ are equivalent. When analysing
  alignment, the orientation of the axis is important, but its direction
  (i.e.\ which end is the `head' and which is the `tail') is arbitrary and
  has no physical meaning. Axial data are synonymous with undirected data,
  unlike vectors which are directed.

Statistical analysis of directed data typically uses vector algebra (known as
Fisher statistics). However, non-directed data cannot be treated as
vectors. \citet{Fisher1993} recommends a statistically valid solution for
axial (2-axial) or, generally, $p$-axial data. First transform the angles to
vector (circular) data as

\begin{ceqn}
\begin{equation} \label{eqpAxialCirc}
    \Theta\ [0^{\circ},\ 360^{\circ}) =
    \begin{cases}
        2 \times \theta \quad \mathrm{for\ axial\ data} & [0^{\circ},\ 180^{\circ})\ ,\\
        p \times \theta \quad \mathrm{for\ }p\mathrm{\text{-}axial\ data} & [0^{\circ},\ 360^{\circ}/p)\ ,
    \end{cases}
\end{equation}
\end{ceqn}

\noindent then analyse the data as required and back-transform the
results. The final step, back-transformation, is generally required only to
find direction \citep{Fisher1993}. In the case of axial data,
back-transformation is simply halving any resultant angles, e.g.\ to
determine the direction of a mean resultant vector.

\citet{Hutsemekers2014} and \citet{Pelgrims2016thesis} test for alignment
and, simultaneously, for orthogonality (which they describe as
`anti-alignment'), by converting from 2-axial $\theta$ [$0^{\circ}$,
  $180^{\circ}$) to 4-axial $\theta_{4ax}$ [$0^{\circ}$, $90^{\circ}$) data
    using

\begin{ceqn}
\begin{equation} \label{eq4Axial}
   \theta_{4ax}\ [0^{\circ},\ 90^{\circ}) = \mathrm{mod}(\theta, 90^{\circ})\ .
\end{equation}
\end{ceqn}

\noindent For vector algebra, this is transformed to circular data $\Theta$
using Eq.~\ref{eqpAxialCirc}, specifically

\begin{ceqn}
\begin{equation} \label{eq4AxialCirc}
   \Theta_{4ax}\ [0^{\circ},\ 360^{\circ}) = 4 \times \theta_{4ax}\ ,
\end{equation}
\end{ceqn}

\noindent where back-transformation, if required, would be to quarter any
resultant angles.

Throughout this work we specify which construct of PA data we use, i.e.\ raw
2-axial ($\theta$), circular 2-axial ($\Theta_{2ax} = 2\theta$,
Eq.~\ref{eqpAxialCirc}), 4-axial ($\theta_{4ax}$, Eq.~\ref{eq4Axial}), or
circular 4-axial ($\Theta_{4ax} = 4\theta_{4ax}$, Eq.~\ref{eq4AxialCirc}).

\section{PA uncertainties}
\label{appPAuncertainties}

To estimate the measurement uncertainties in the LQG position angles we use
bootstrap re-sampling with replacement \citep{Efron1979}. For an LQG with $m$
members, the dataset consists of $m$ observed quasar positions (right
ascension, declination, and redshift). We construct $n$ bootstrap LQGs, each
with the same number of $m$ members which are drawn at random from the
original dataset. Each draw is made from the entire dataset, and each member
is replaced in the dataset before the next draw. Thus, each bootstrap LQG is
likely to miss some members and have duplicates (or triplicates or more) of
others.

These bootstraps are used to estimate the uncertainty on parameters
(e.g.\ PAs) derived from the dataset, without any assumption about the
underlying population \citep{Feigelson2012}. The position angle of each
bootstrap LQG is calculated through the same 2D and 3D approaches used for
the observed LQGs (Appendix~\ref{appLQGorientation}).

For a sample of $n$ bootstrap LQGs we can determine the mean PA and its
associated uncertainty. Using the linear mean $\bar{\theta} = (\sum_{i =
  1}^{n}\theta_i) / n$ is inappropriate for axial data [$0^{\circ}$,
  $180^{\circ}$), where $0^{\circ}$ and $180^{\circ}$ are equivalent. For
  example, a sample of PAs centred on $0^{\circ}$, with around half in the
  range $0^{\circ} \lesssim \theta \lesssim 10^{\circ}$ and half in the range
  $170^{\circ} \lesssim \theta \lesssim 180^{\circ}$, has a linear mean
  $\bar{\theta} \sim 90^{\circ}$ rather than the correct answer $\bar{\theta}
  \sim 0^{\circ}$ (or, equivalently, $\bar{\theta} \sim 180^{\circ}$).

Therefore, instead of linear mean, we calculate the `circular' mean
\citep{Fisher1993} of axial PAs. First, following \citet{Mardia2000}, let

\begin{ceqn}
\begin{equation} \label{eqCosSin}
    \bar{C} = \frac{1}{n} \sum_{i = 1}^{n} \cos 2\theta_i\ , \ \ \bar{S} = \frac{1}{n} \sum_{i = 1}^{n} \sin 2\theta_i\ ,
\end{equation}
\end{ceqn}

\noindent where $\theta_i$ is the PA of the $i^{\mathrm{th}}$ bootstrap LQG
and $n$ is the total number of boostraps for this LQG. The factor of two
accounts for the axial (rather than circular) nature of the data. The mean
direction $\bar{\theta}$ is given by

\begin{ceqn}
\begin{equation} \label{eqCircMean}
    \bar{\theta} =
    \begin{cases}
        \dfrac{1}{2} \arctan \left( \dfrac{\bar{S}}{\bar{C}} \right) & \mathrm{if}\ \bar{C} \geq 0\ ,\\
        \dfrac{1}{2} \arctan \left( \dfrac{\bar{S}}{\bar{C}} \right) + \pi & \mathrm{if}\ \bar{C} < 0\ ,
    \end{cases}
\end{equation}
\end{ceqn}

\noindent where the factor $\frac{1}{2}$ converts from vector algebra back to
axial data (i.e.\ `back-transformation').

\begin{figure} 
    \centering
    \subfigure[Example 1 - narrow LQG]{\includegraphics[width=0.6\columnwidth]{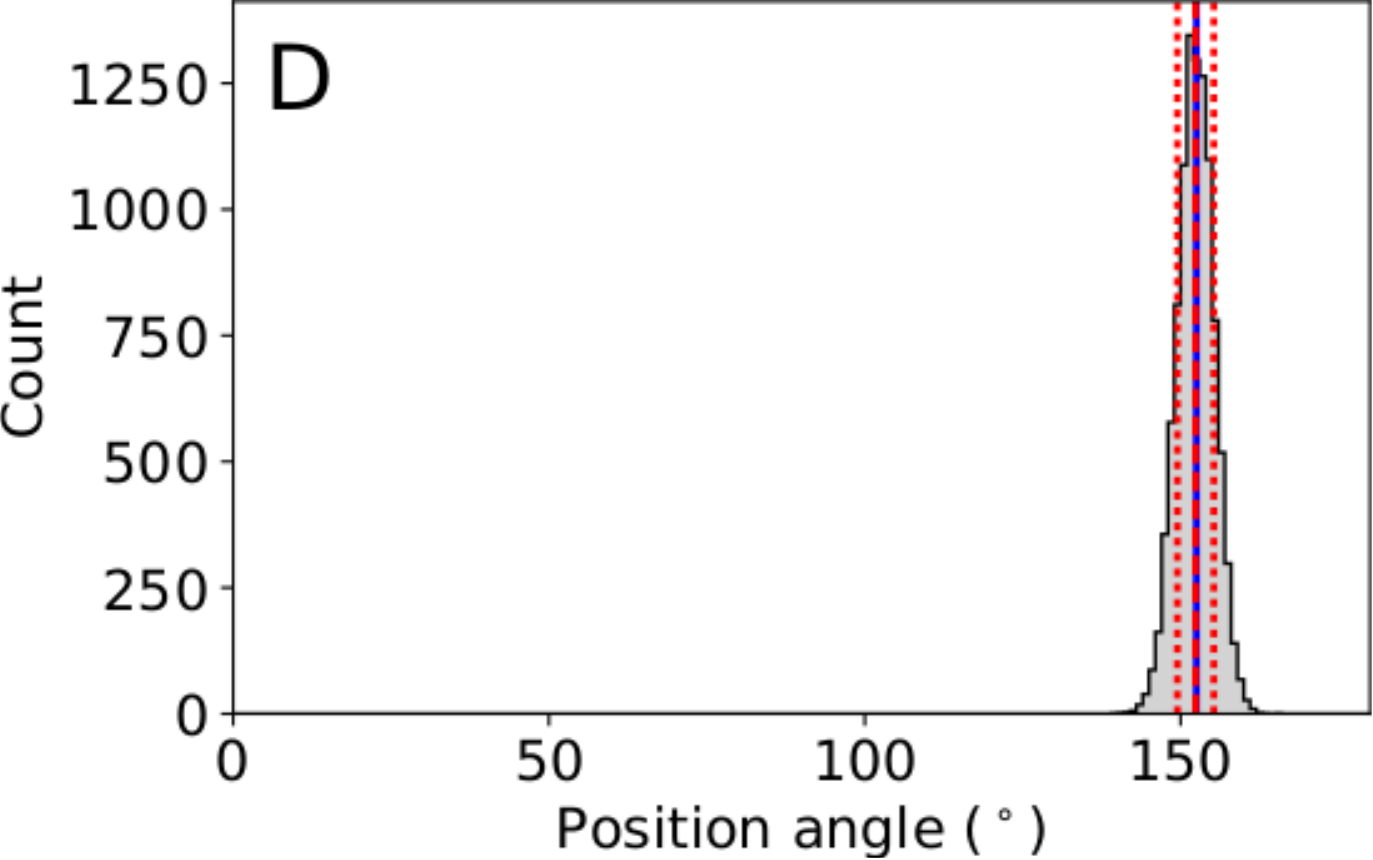} \label{subBSNarrow}}\\ 
    \subfigure[Example 2 - broad LQG]{\includegraphics[width=0.6\columnwidth]{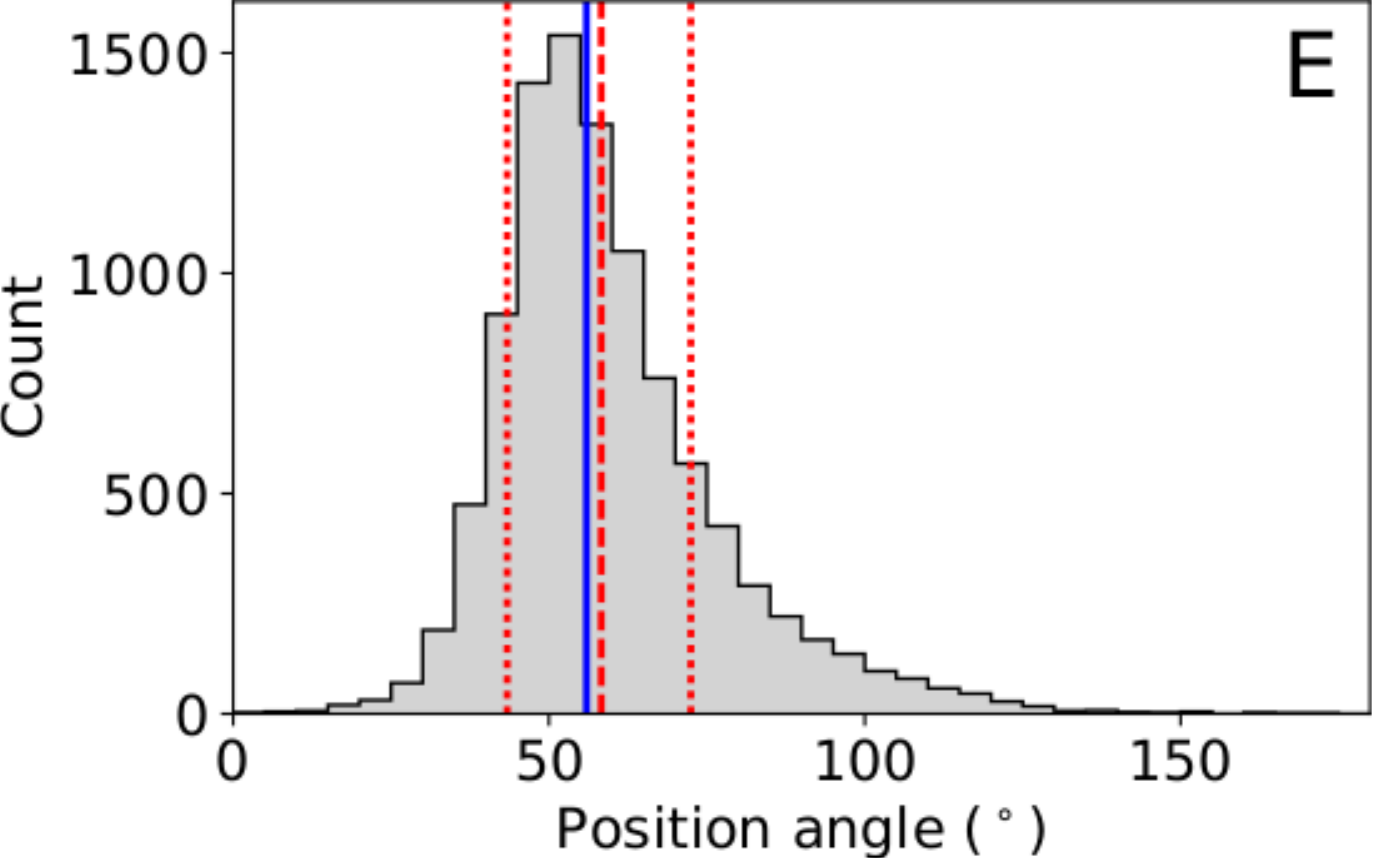} \label{subBSWide}}\\ 
    \subfigure[Example 3 - intermediate LQG]{\includegraphics[width=0.6\columnwidth]{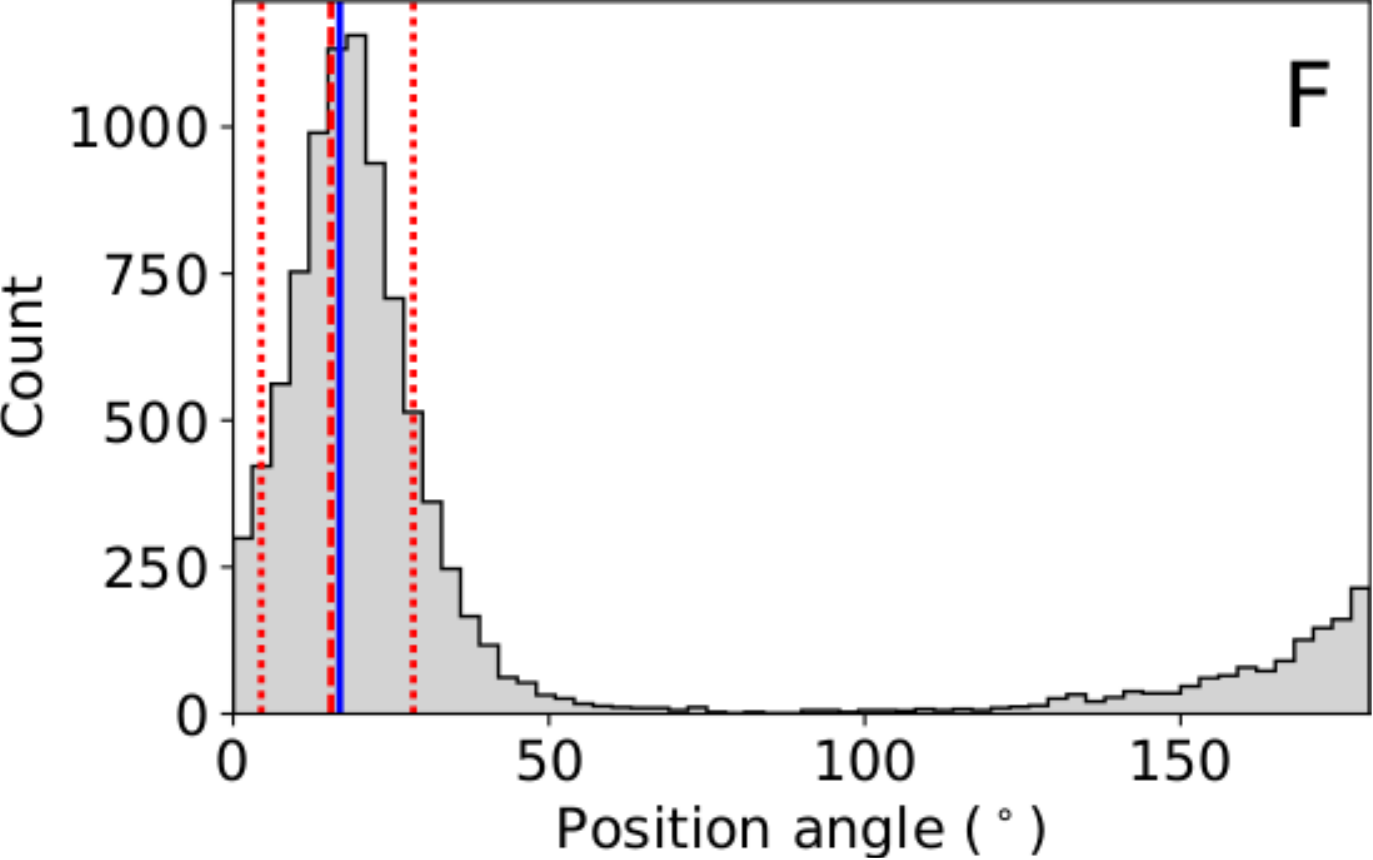} \label{subBSWrap}} 
    \caption[Position angle distributions from bootstraps of three example
      LQGs]{LQG position angles from 10000 bootstraps for three example LQGs,
      labelled D, E, F in Fig.~\ref{figAllLQGsPt1}, and visually classified
      as \protect\subref{subBSNarrow} narrow, \protect\subref{subBSWide}
      broad, and \protect\subref{subBSWrap} intermediate morphology. All PA
      calculations use the 2D approach with no parallel transport. Dashed
      orange line shows the circular mean of the bootstraps $\bar{\theta}$,
      dotted orange lines show the 68\% confidence interval, solid blue line
      shows the observed PA.}
    \label{figBootstraps}
\end{figure}

For each LQG, to estimate the uncertainty in the mean $\bar{\theta}$, we calculate its confidence interval following \citet{Pelgrims2016thesis}, who in turn follows \cite{Fisher1993}, and for each individual bootstrap $i$ out of a total of $n$ we define the residual as

\begin{ceqn}
\begin{equation} \label{eqCircGammas}
    \gamma_i = \frac{1}{2} \arctan \left( \frac{\sin(2(\theta_i - \bar{\theta}))}{\cos(2(\theta_i - \bar{\theta}))}\right)\ ,
\end{equation}
\end{ceqn}

\noindent where $\theta_i$ is the PA of the $i^{\mathrm{th}}$ bootstrap
LQG. For $i = 1,...,n$ we sort the $\gamma_i$ in ascending order to give an
ordered list $\gamma_{(1)} \leq ... \leq \gamma_{(n)}$. To determine the
confidence interval at the $100(1 - \alpha)\%$ level\footnote{Confidence
  level $= 1 - \alpha$, where $\alpha$ is the significance level} we find the
$\gamma_i$ list elements at lower index $l$ which is the integer part of
$(n\alpha + 1) / 2$ and upper index $u = n - l$. The confidence interval for
$\bar{\theta}$ is then $[\bar{\theta} + \gamma_{(l + 1)}, \bar{\theta} +
  \gamma_{(u)}]$. We calculate the confidence interval at the 68\% level
($\alpha = 0.32$) and define the half-width of the confidence interval (HWCI)
as

\begin{ceqn}
\begin{equation} \label{eqHWCI}
    \gamma_h = (\gamma_{(u)} - \gamma_{(l + 1)}) / 2 \ .
\end{equation}
\end{ceqn}

For each of our sample of 71 LQGs, we create $n = 10000$ bootstraps,
calculate their PAs $\theta_i$ using the 2D or 3D approach
(Appendix~\ref{appLQGorientation}), and then determine the mean
$\bar{\theta}$ and confidence interval. Fig.~\ref{figBootstraps} shows the
distribution of bootstrap PAs $\theta_i$ calculated using the 2D approach for
three example LQGs. The circular mean of the bootstraps $\bar{\theta}$
(dashed orange line) generally agrees well with the observed PA (solid blue
line). As expected, the 68\% confidence interval (Fig.~\ref{figBootstraps},
dotted orange lines) is smaller for (a) a `narrow' LQG than (b) a `broad'
sheet-like LQG. Example (c) illustrates why the linear mean is inappropriate,
since the distribution of the bootstrap PAs may `wrap' from $180^{\circ}$
back to $0^{\circ}$.

The HWCIs for 10000 bootstraps of our sample of 71 LQGs, with PAs $\theta_i$
calculated using both the 2D and 3D approaches, are shown as error bars in
Fig.~\ref{figPA2D3D}. The mean (median) HWCI for the 2D approach is
$\sim10^{\circ}$ ($\sim8^{\circ}$), and for the 3D approach it is
$\sim11^{\circ}$ ($\sim9^{\circ}$).

\section{Applying the S test}
\label{appSTest}

\subsection{Nearest neighbours free parameter}
\label{subSTestNeighbours}

The S~test quantifies the coherence of PA alignment by measuring the
dispersion of groups of $n_v$ nearest neighbours, where $n_v$ is a free
parameter. We explore a range of $n_v$ values; we do not choose a specific
value. For each LQG, its nearest neighbours can be determined either in two
dimensions (angular separation) or three dimensions (proper separation).

In 2D, nearest neighbours are identified by calculating the angular
separation $\theta$ on the celestial sphere between LQG 1 and LQG 2 as

\begin{ceqn}
\begin{equation}
    \theta = \cos^{-1}[\sin \delta_1 \sin \delta_2 + \cos \delta_1 \cos \delta_2 \cos(\alpha_1 - \alpha_2)] \ ,
\end{equation}
\end{ceqn}

\noindent where $\alpha_1$ and $\delta_1$ ($\alpha_2$ and $\delta_2$) are the
right ascension and declination of the centroids of LQG 1 (2)
respectively. For each LQG, its nearest neighbours are those $n_v$ LQGs
separated from it by the smallest angular distances.

In 3D, nearest neighbours are identified by calculating the three-dimensional
proper positions ($x$, $y$, $z$) of each LQG centroid. For each LQG, its
nearest neighbours are those $n_v$ LQGs separated from it by the smallest
proper distances.

The 2D and 3D approaches of identifying nearest neighbours will return
different groups of LQGs. When these groups are used to compute the S
statistic $S_D$ we find that the values calculated using the two approaches
are remarkably consistent. This is probably due to the geometry of the
three-dimensional survey volume; our redshift restriction of $1.0 \le z \le
1.8$ yields a `shell' of LQGs of finite thickness. At low $n_v$ the 3D
approach may find neighbours in the radial direction, but at high $n_v$ it
can only find them tangentially (on the sky), like the 2D approach. Since the
two approaches give very similar results, and the 3D approach is more
physically motivated, we use the 3D approach of identifying nearest
neighbours in this work.

\subsection{Estimating significance level}
\label{subSTestRandoms}

The S~test yields an $S_D$ statistic for each number of nearest neighbours
$n_v$ assessed. The significance level of these values compared to randomness
cannot be evaluated analytically, due to: overlaps between groups of nearest
neighbours, deviation of the $S_D$ distribution from normality (particularly
for small $n$ and when $n_v \sim n$), and the dependence of parallel
transport corrections on the precise location of the objects
involved. Therefore, numerical simulations are required.

\citet{Hutsemekers1998} and \citet{Hutsemekers2001} generate random samples
by shuffling the observed PAs randomly between objects, while keeping their
positions fixed. This has the effect of erasing any correlation between PAs
and positions. But, as \citet{Jain2004} note, the shuffling method is unable
to test for global alignment.

We have no reason to expect LQG PAs to be correlated with LQG positions;
alignment could be global. Therefore, we generate random samples from a
uniform distribution, keeping the three-dimensional LQG positions fixed. We
generate $n = 71$ PAs, randomly drawn from a uniform distribution. These are
generated in the ranges [$0^{\circ}$, $180^{\circ}$) for 2-axial PAs and
  [$0^{\circ}$, $90^{\circ}$) for 4-axial PAs. Note that initially generating
    4-axial PAs in the range [$0^{\circ}$, $180^{\circ}$), then converting to
      [$0^{\circ}$, $90^{\circ}$) after parallel transport and before
        evaluating the S~test is equivalent.

The significance level (SL) of the S~test is defined as the percentage of
simulations that have an $S_D$ statistic at least as extreme as the one from
our observations \citep{Pelgrims2016thesis}. It is computed by comparing the
statistic of our observations ($S_{D, obs}$) with the statistic from a large
number of numerical simulations ($S_{D, sim}$).

For 2-axial PAs we find that $S_{D, obs}$ is in the low $S_{D, sim}$
(left-hand) tail of the distribution. To interpret this, consider the
potential bimodality of the LQG PA distribution, with the peaks separated by
$\sim 90^{\circ}$. This orthogonality leads to a large dispersion, and yields
a small value of $S_D$. Therefore, it is legitimate to interpret a result in
the left-hand tail as a potentially orthogonal signal, with the SL being the
probability that $S_{D, sim} < S_{D, obs}$. This can be tested by also
evaluating the S~test for 4-axial PAs.

Note that two modes of a possible combined `alignment plus orthogonality'
signal would tend to erase any signal, reducing the power of this
test. However, a residual signal in the left (right) tail indicates the
orthogonality (alignment) mode dominates.

When we convert 2-axial PAs to 4-axial, we simultaneously test for alignment
and orthogonality by combining the modes, so an `alignment plus
orthogonality' signal will manifest as alignment only. The SL is therefore
the proportion of numerical simulations with statistic $S_D$ higher than that
observed. Indeed, for 4-axial PAs we find that $S_{D, obs}$ is in the high
$S_{D, sim}$ (right-hand) tail of the distribution, with the SL being the
probability that $S_{D, sim} > S_{D, obs}$. Note that an `alignment only'
signal would also be in the right-hand tail, but would be differentiated by
its 2-axial result.

The $S_D$ distributions of 2-axial and 4-axial numerical simulations differ,
particularly as $n_v$ approaches $n$. Therefore, we generate separate
numerical simulations for each, and calculate the significance levels as

\begin{ceqn}
\begin{equation} \label{eqSL2ax}
    \mathrm{SL_{2ax}} = P(S_{D, sim(2ax)} < S_{D, obs(2ax)})\ ,
\end{equation}
\end{ceqn}

\begin{ceqn}
\begin{equation} \label{eqSL4ax}
    \mathrm{SL_{4ax}} = P(S_{D, sim(4ax)} > S_{D, obs(4ax)})\ ,
\end{equation}
\end{ceqn}

\noindent where $P$ indicates probability, $2ax$ and $4ax$ indicate 2-axial
and 4-axial PAs, and $sim$ and $obs$ indicate simulations and observations
respectively.

\section{LQGs in 3D proper space}
\label{appLQGs3DPerspectives}

\begin{figure} 
    \centering
    \includegraphics[width=0.9\columnwidth]{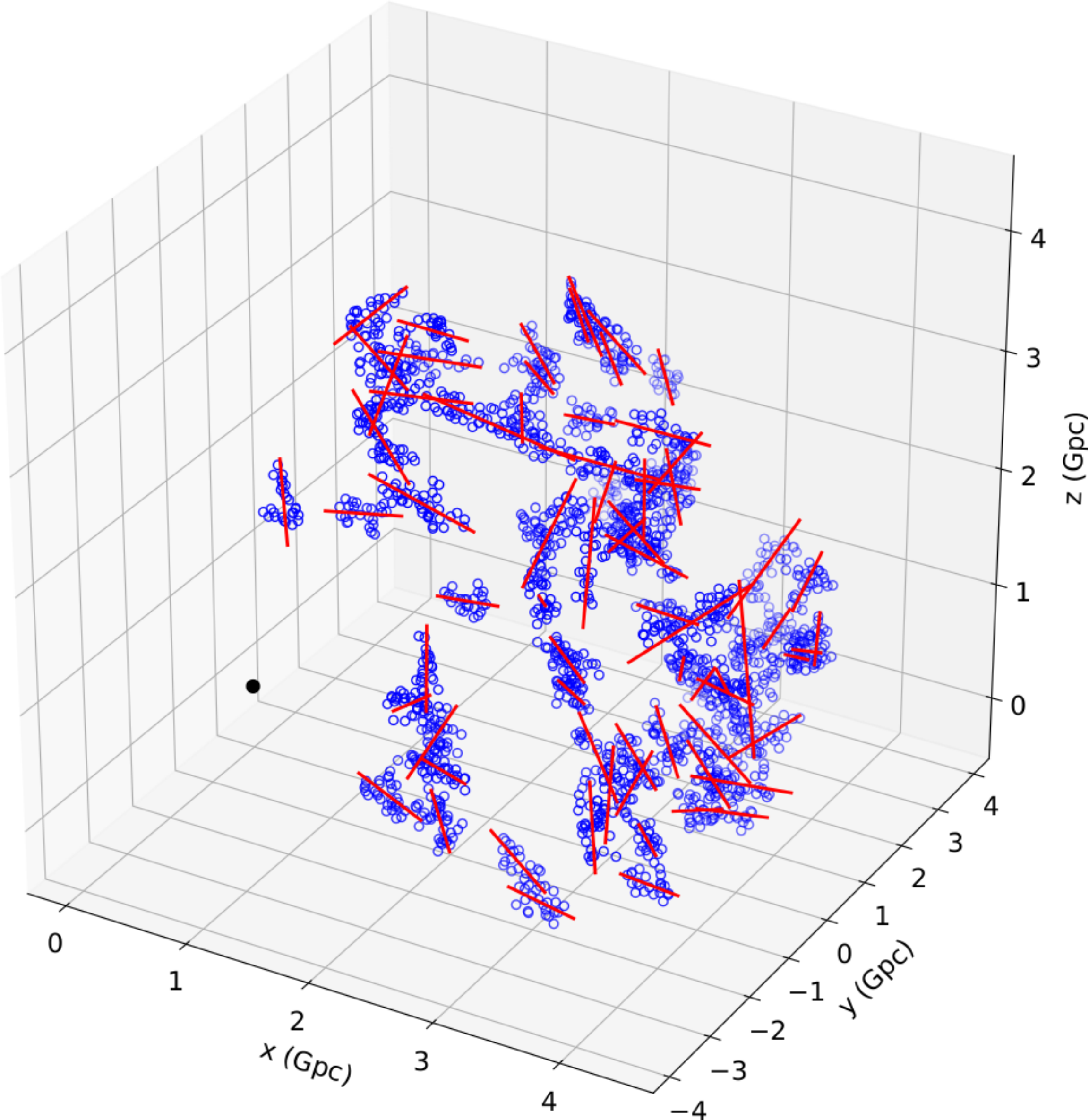}
    \caption[LQGs and their major axes in three-dimensional proper
      space]{LQGs in three-dimensional proper space, showing LQG quasars
      (blue circles) and LQG major axes (red lines). Quasar markers are
      shaded to give the appearance of depth, with lighter shades
      representing more distant quasars. Our location at ($x, y, z$)
      coordinates (0, 0, 0) is indicated by a black dot. See
      Fig.~\ref{figLQGs3DAspects} for three alternative, orthogonal
      perspectives.}
    \label{figLQGs3D}
\end{figure}

\begin{figure} 
    \centering
    \subfigure[$x$-axis orthogonal to the page]{\includegraphics[width=0.85\columnwidth]{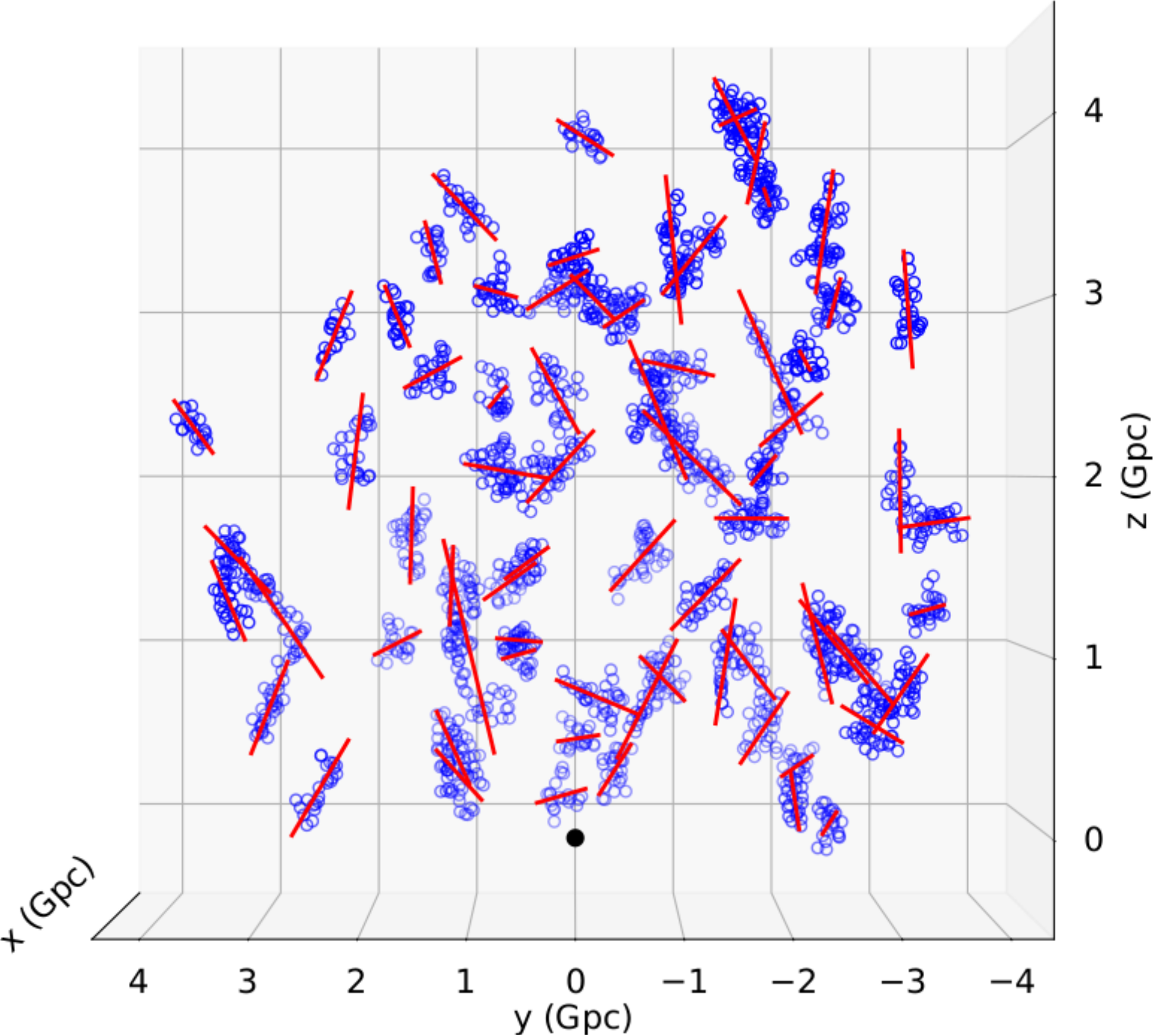} \label{subxaxes}}\\
    \subfigure[$y$-axis orthogonal to the page]{\includegraphics[width=0.85\columnwidth]{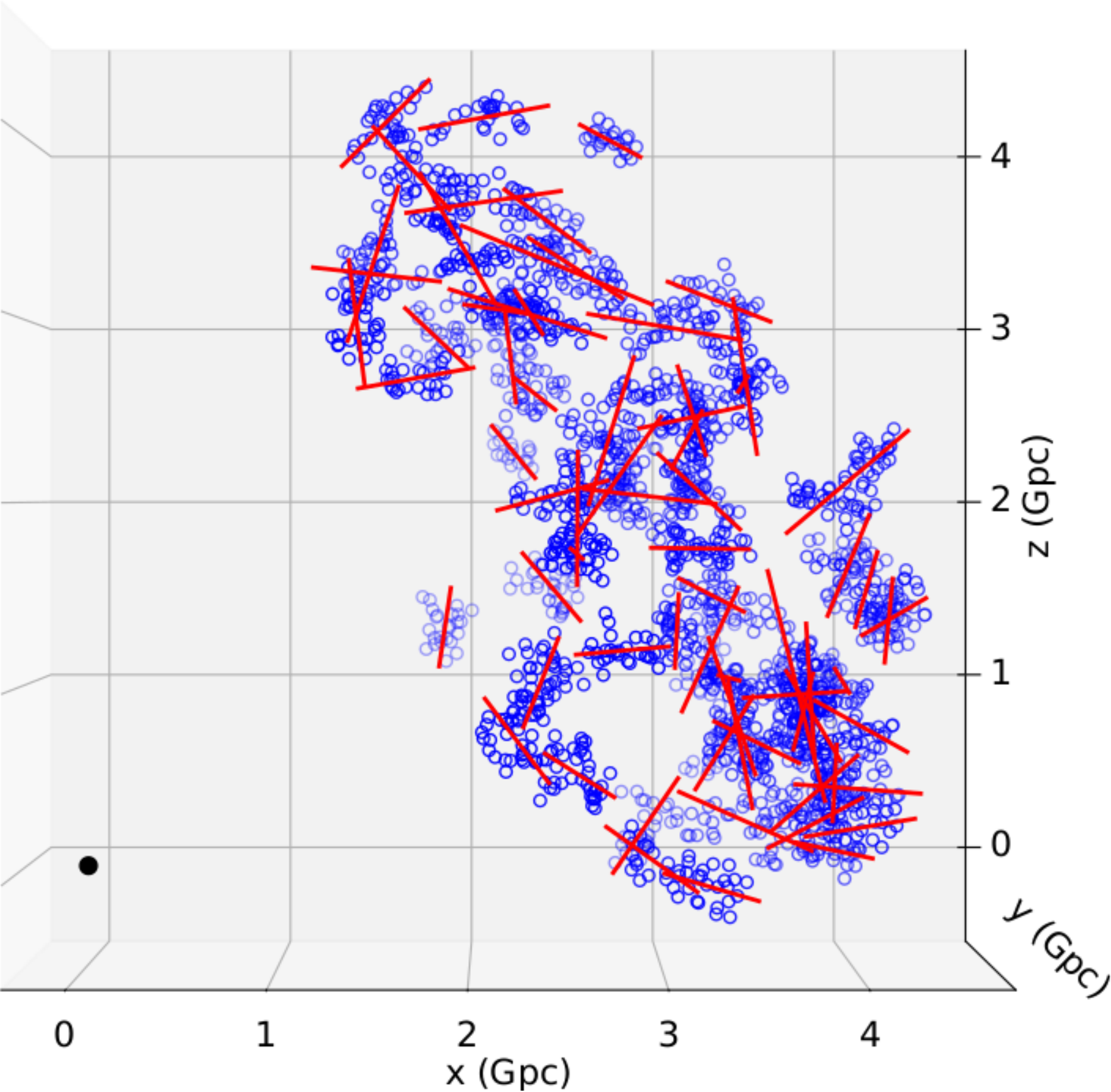} \label{subyaxis}}\\
    \subfigure[$z$-axis orthogonal to the page]{\includegraphics[width=0.85\columnwidth]{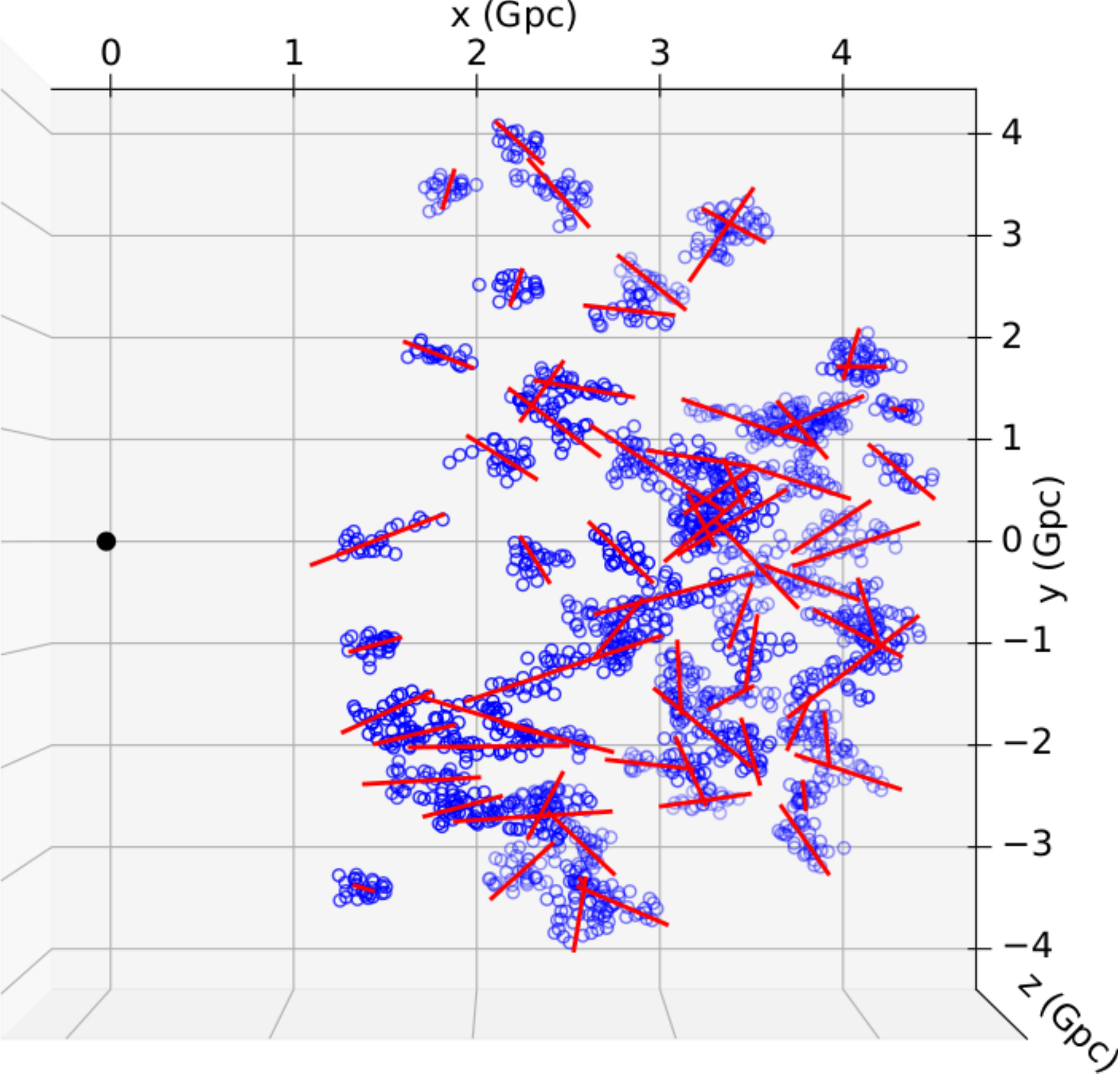} \label{subzaxis}}
    \caption[As Fig.~\ref{figLQGs3D} but viewed from three orthogonal
      perspectives]{As Fig.~\ref{figLQGs3D} but viewed with the (a) $x$-axis,
      (b) $y$-axis, and (c) $z$-axis orthogonal to the page.}
    \label{figLQGs3DAspects}
\end{figure}

The 71 LQG positions in three-dimensional proper space, and their
orientation as determined by the three-dimensional method, are illustrated in
Fig.~\ref{figLQGs3D}, with additional perspectives shown in
Fig.~\ref{figLQGs3DAspects}.

\section{Parallel transport destination}
\label{appGaussians}

\begin{figure*} 
    \centering
	\includegraphics[width=1.9\columnwidth]{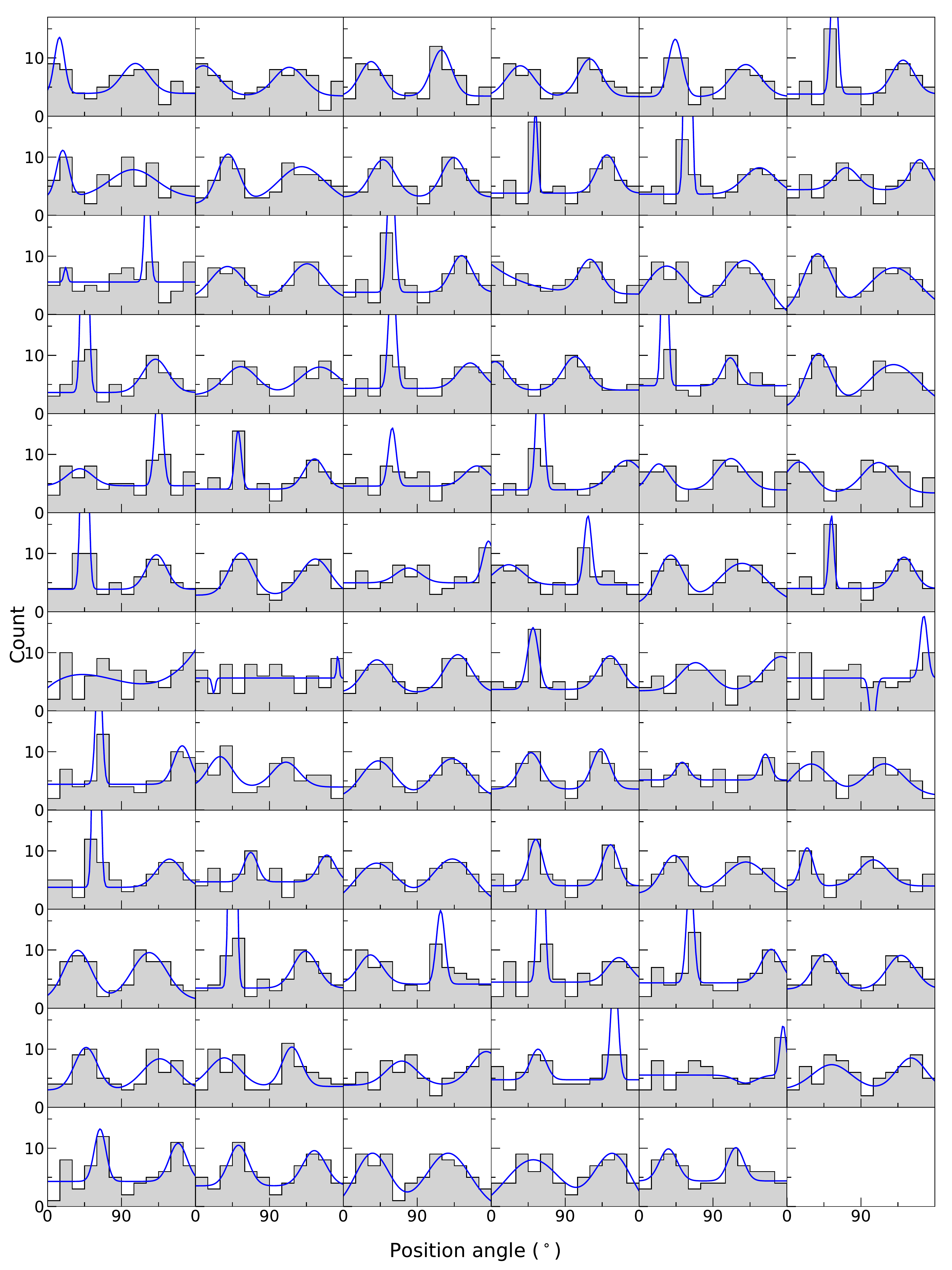}
	\caption[Position angles parallel transported to each LQG
          position]{LQG position angles after parallel transport to the
          location of each LQG; $15^{\circ}$ bins. Solid blue lines are
          double Gaussian fit. The bimodal distribution is generally robust
          to parallel transport destination.}
	\label{figGaussians}
\end{figure*}

Fig.~\ref{figGaussians} shows LQG PAs after parallel transport to the
location of each of the 71 LQGs. Also shown, double Gaussian fit (blue).

\bsp	
\label{lastpage}
\end{document}